\documentclass{article}

\usepackage{microtype}
\usepackage{graphicx}
\usepackage{subcaption}
\usepackage{booktabs} % for professional tables

\usepackage{hyperref}

\usepackage[preprint]{icml2026}

\usepackage{amsmath}
\usepackage{amssymb}
\usepackage{mathtools}
\usepackage{amsthm}

\usepackage[capitalize,noabbrev]{cleveref}

\theoremstyle{plain}

\theoremstyle{definition}

\theoremstyle{remark}

\usepackage[textsize=tiny]{todonotes}

\usepackage{makecell}

\icmltitlerunning{NIC-RobustBench: Benchmark for Neural Image Compression and Robustness Analysis}

\begin{document}

\twocolumn[
  \icmltitle{NIC-RobustBench: A Comprehensive Open-Source Toolkit and Benchmark for Neural Image Compression and Robustness Analysis}



  \icmlsetsymbol{equal}{*}

  \begin{icmlauthorlist}
    \icmlauthor{Georgii Bychkov}{isp,iai}
    \icmlauthor{Khaled Abud}{isp,iai,msu}
    \icmlauthor{Egor Kovalev}{iai,msu}
    \icmlauthor{Aleksandr Gushchin}{isp,iai,msu}
    \icmlauthor{Sergey Lavrushkin}{isp,iai}
    \icmlauthor{Dmitriy S. Vatolin}{isp,iai,msu}
    \icmlauthor{Anastasia Antsiferova}{isp,iai,inno}
  \end{icmlauthorlist}

  \icmlaffiliation{iai}{MSU Institute for Artificial Intelligence, Moscow, Russia}
\icmlaffiliation{msu}{Lomonosov Moscow State University, Moscow, Russia}
\icmlaffiliation{isp}{ISP RAS Research Center for Trusted Artificial Intelligence, Moscow, Russia}
\icmlaffiliation{inno}{Innopolis University, Innopolis, Russia}
%\icmlaffiliation{aimsu}{AI Center, Lomonosov Moscow State University, Moscow, Russia}

  \icmlcorrespondingauthor{Georgii Bychkov}{georgy.bychkov@graphics.cs.msu.ru}

  % paper; these are used to populate the "keywords" metadata in the PDF, but
  \icmlkeywords{Machine Learning, ICML}

  \vskip 0.3in
]



\printAffiliationsAndNotice{}  % no special notice (required even if empty)

\begin{abstract}
  Neural image compression (NIC) is increasingly used in computer vision pipelines, as learning-based models are able to surpass traditional algorithms in compression efficiency. However, learned codecs can be unstable and vulnerable to adversarial attacks: small perturbations may cause severe reconstruction artifacts or indirectly break downstream models. Despite these risks, most NIC benchmarks only emphasize rate-distortion (RD) performance, focusing on model efficiency in safe, non-adversarial scenarios, while NIC robustness studies cover only specific codecs and attacks. To fill this gap, we introduce \textbf{NIC-RobustBench}, an open-source benchmark and evaluation framework for adversarial robustness of NIC methods. The benchmark integrates 8 attacks, 9 defense strategies, standard RD metrics, a large and extensible set of codecs, and tools for assessing both the robustness of the compression model and impact on downstream tasks. Using NIC-RobustBench, we provide a broad empirical study of modern NICs and defenses in adversarial scenarios, highlighting failure modes, least and most resilient architectures, and other insights into NIC robustness. Code is made available at \textit{[link]}.
\end{abstract}

\section{Introduction}
% сокращенная версия:
Effective image compression is fundamental to digital media, enabling efficient storage and transmission of vast volumes of visual data generated in modern applications. Compression remains crucial for reducing data storage requirements and network load. Traditionally, image codecs rely on hand-crafted transforms and heuristics; however, the advances in deep learning have prompted the development of neural-network-based methods \cite{nic1, nic2, nic3}. In recent years, researchers have proposed numerous neural image compression (NIC) models that use learned modules to replace or augment components of the classical pipeline. These approaches have achieved state-of-the-art performance, culminating in the first neural image compression standard, JPEG AI \cite{jpeg_ai_standard}.

\begin{figure}[tb]
    \centering
    \includegraphics[width=.95\linewidth]{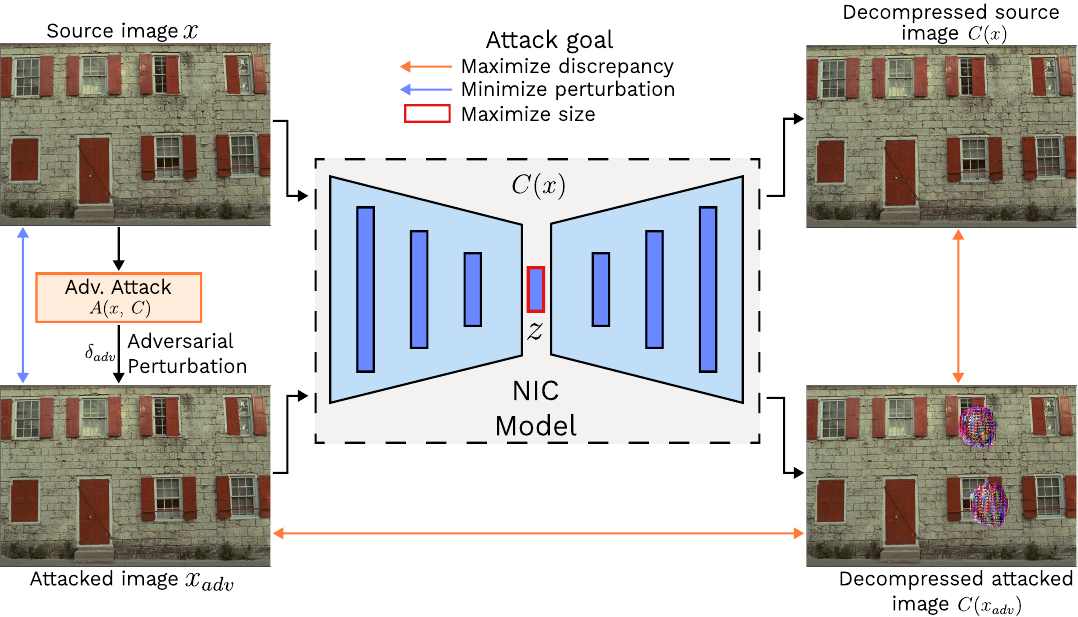}
    \caption{Adversarial attack on NIC with possible attack goals.}
    \label{fig:att_process}
\end{figure}

However, NIC models inherit the susceptibility of neural networks to adversarial attacks. An adversarial attack involves adding a carefully crafted perturbation to an input that causes a model to produce incorrect or highly degraded outputs \cite{dong2018boosting, chakraborty2021survey}. Adversarial robustness is a well-established concern in computer vision tasks such as classification \cite{croce2020robustbench}, detection \cite{nezami2021pick}, and quality assessment~\cite{iqarobustness}, yet it remains underexplored in learned image compression. As illustrated in Fig.~\ref{fig:att_process}, adding subtle, carefully crafted perturbations to an input can cause NIC models to produce severe reconstruction errors, significantly degrading output quality.

\begin{figure*}[t!]
\centering
\includegraphics[width=0.95\textwidth]{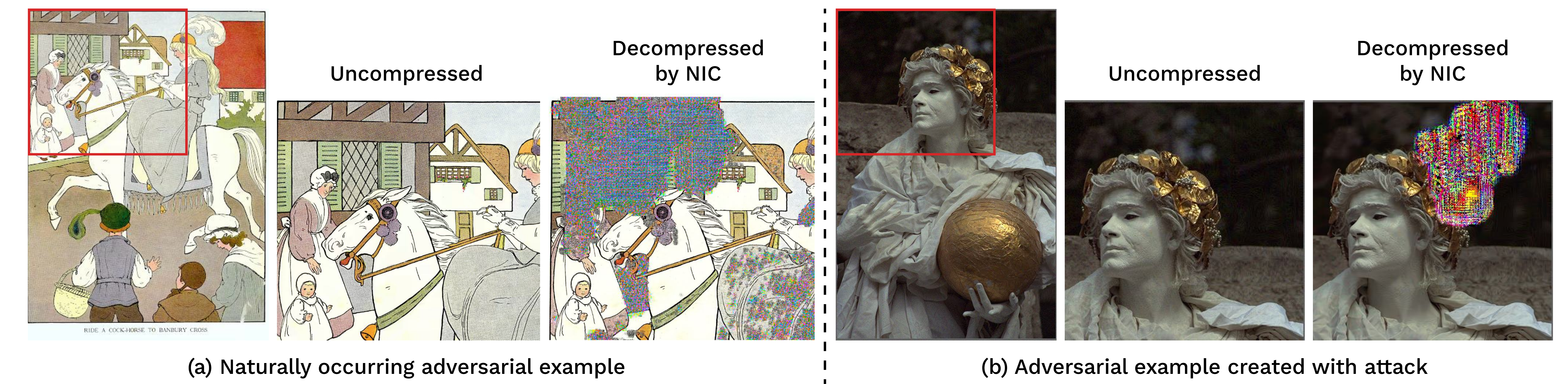}
\caption{Examples of natural and artificial adversarial examples for neural compression (JPEG AI model). The left image is taken from the Open Images dataset \cite{kuznetsova2020open}, the right is taken from the Kodak dataset \cite{kodakDataset}.}
\label{fig:art_axamples}
\end{figure*}

% сокращенная версия:
Assessing NIC robustness is crucial for two key reasons. First, learned codecs can exhibit unstable behavior: even without malicious intent, they can produce severe, non‑traditional artifacts on certain natural images, as recently demonstrated with JPEG AI \cite{tsereh2024jpegai}. As illustrated in Fig~\ref{fig:art_axamples}, some of these failure cases resemble adversarial examples, indicating intrinsic fragility in current NIC approaches. Second, in real-world scenarios, compression often serves as a preprocessing step for downstream computer vision tasks. An adversary can target the NIC stage to introduce corruptions that indirectly compromise subsequent models (e.g., object detectors), undermining the whole pipeline. Therefore, evaluating and enhancing the adversarial robustness of NIC systems is vital both for reliable compression and for protecting downstream systems. Evaluating NIC robustness poses unique challenges compared to other tasks: adversarial perturbations can alter both reconstruction quality and bitrate, necessitating robustness benchmarks to apply different approaches than those used in classification or detection tasks. 

Existing open-source NIC libraries such as CompressAI ~\cite{begaint2020compressai}, OpenDIC ~\cite{gao2024opendic}, and TFC~\cite{tfc_github} primarily evaluate rate-distortion performance on clean data. While several studies have investigated the robustness of individual NIC models and proposed specific adversarial attacks, none have consolidated these efforts into a unified framework that supports adversarial attacks, robustness metrics, defenses, and downstream-task evaluation across modern NIC architectures.

%Our work introduces novel attack strategies by using new attack objectives that were not explored in previous work, notably targeting bitrate.

% сокращенная версия:
To bridge this gap, we propose \textbf{NIC-RobustBench}, the first open-source toolkit designed to benchmark and improve the adversarial robustness of NIC models. Our goal is to provide a standardized, extensible framework that enables consistent comparison across NIC models, adversarial attacks, and defenses. Implemented in a modular manner for easy scalability, our framework enables researchers to systematically evaluate a broad range of algorithms under diverse adversarial scenarios and assess various defensive strategies for NIC models. Leveraging our framework, we conduct a series of large-scale evaluations and establish a comprehensive NIC robustness benchmark that covers a diverse set of codecs, attacks, and defense mechanisms.

Our main contributions are summarized as follows:

\begin{itemize}
\item \textbf{First large-scale NIC robustness benchmark}: We establish the first comprehensive adversarial robustness benchmark for neural image compression. NIC-RobustBench facilitates rigorous comparative analysis by covering the largest collection of NIC models among available libraries, including the latest JPEG AI codec.
\item \textbf{Scalable framework}: We provide a modular, open-source library that supports multiple types of attacks, adversarial objectives, and defenses for the learned image compression task. Designed for easy scalability, NIC-RobustBench allows seamless integration of new components and evaluation of attacks' impact on both image quality and downstream tasks. 
\item \textbf{Comprehensive evaluation}: We perform extensive experiments with 10 diverse NIC models across 5 datasets, using 8 different attacks with 6 unique attack objectives targeting different aspects of NIC performance. Our evaluation provides valuable insights into the robustness of different NIC architectures in novel adversarial scenarios.
\item \textbf{Defense strategies}: We implement and investigate multiple strategies to defend NIC models against adversarial attacks, and identify the most effective techniques at improving the NIC robustness of neural image compressors.
\end{itemize}

\section{Related Work}
\label{gen_inst}

%\subsection{Neural image compression}
\label{sec:nic_models}

\textbf{Adversarial robustness of NIC.} \cite{malic} proposed a bitrate-oriented adversarial attack on NIC models based on I-FGSM \cite{ifgsm}. Their study explored the impact of this attack across multiple codec architectures, highlighting design components that contribute to enhanced robustness, and identified their proposed factorized attention model as the most stable architecture. %Chen and Ma
Similarly, \cite{chen2023toward} demonstrated that NIC methods are highly susceptible to adversarial perturbations that reduce the quality of the decoded image. They adopted the I-FGSM and C\&W \cite{cw} attacks to increase the difference between the compressed images before and after the attack and introduced the Fast Threshold-constrained Distortion Attack (FTDA) that attacks images after compression. Furthermore, they proposed the $\Delta \text{PSNR}$ metric for evaluating attack performance, which jointly captures the trade-off between attack effectiveness and perceptual visibility. \cite{song2024training} defends models using randomized image transformations and a two-way compression strategy. \cite{kalmykov2026t} create adversarial perturbations in the wavelet domain to improve the stealthiness of the attack.

Sec. \ref{sec:related_works} discusses related work on NIC models.

%\cite{song2024training} proposed to 
% They also proposed a $\Delta \text{PSNR}$ method for assessing the success of an attack, which evaluates both its effectiveness and its visibility: 
% \begin{equation*} 
% \Delta \text{PSNR} = \text{PSNR}(x, C(x)) - \text{PSNR}(x', C(x')),
% \end{equation*}
% where $x$ is the original image, $x'$ is the adversarial image, $C(\cdot)$ - image after NIC.

\section{Problem Formulation}
\textbf{NIC models.}
Lossy image compression is based on a rate-distortion theory. The goal is to find a trade-off between the size of the compressed image representation and the decrease in the perceptual quality of a reconstructed image: %This problem is formulated as follows:
\begin{equation}
     \theta= \underset{\theta}{\arg\min} [\lambda r(\hat{y}(\theta)) + d(x,\hat{x}(\theta))],
\end{equation}
where $\theta$ means weights of the NIC model, $r(\hat{y}(\theta))$ is a  bitrate of a quantized image, and $d(x,\hat{x}(\theta))$ --- perceptual image-similarity metric. The encoder-decoder architecture is one of the possible solutions. For a given image $x \in X = \mathbb{R}^{H\times W \times 3}$, an encoder $E$ transforms it to a latent representation $y=E(x)$. Then, the data is quantized $\hat{y}=Q(y)$, and a decoder $G$ performs the reconstruction of the image $\hat{x}=G(\hat{y})$. We denote $C(x)$ as a complete encoding-decoding process $C(\cdot) = G \circ Q \circ E: X \to X$.

\textbf{Adversarial attacks for NIC.}
An attack seeks a perturbation $\delta$ that, when added to the original image, makes the adversarial image $x' = x + \delta$ such that its decoded image $C(x')$ differs from the original image as much as possible. Adversarial attack $A: X \rightarrow X$ is defined as follows:
\begin{equation}
    \begin{aligned}
    A(x)= \underset{x': \rho(x',x) \le \varepsilon} {\arg\max} \: L(x, x', C(x), C(x')),
    \end{aligned}
\end{equation}
where $\rho(x',x) = \|\delta\|$, $\varepsilon$ imposes a constraint on the perturbation magnitude, $L: X \times X \to \mathbb{R}$ is a corresponding optimization target. To achieve this goal, we consider 6 loss functions for all employed attacks. They reflect different approaches to measuring the distance between the original and adversarial images and their reconstructed versions. Additionally, we consider an alternative optimization goal. Instead of increasing the distance between the image targets, we reduce the compression ratio of the NIC measured in Bits Per Pixel (BPP):
\begin{equation}
    \begin{aligned}
    A(x)= \underset{x': \rho(x',x) \le \varepsilon} {\arg\max} \: \text{BPP}(Q(E(x'))).
    \end{aligned}
\end{equation}

\iffalse

\begin{table}[tb]
% \floatsetup{floatrowsep=qquad, captionskip=4pt}
% \begin{floatrow}

% \caption{Global caption}
\centering
\begin{tabular}{ll}
Attack & Description\\
\midrule
FTDA \cite{chen2023toward} & \makecell[l]{NIC attack to increase distance \\ between decoded images}\\
I-FGSM \cite{ifgsm} & Iterative sign gradient descent \\
MADC \cite{madc} & Proj. grad. on a proxy metric (MSE)\\
PGD \cite{pgd} & I-FGSM with random initialization \\
SSAH \cite{ssah} & Grad. desc. in high freq. domain \\
CAdv \cite{cadv} & Gradient descent with color filter \\
Random noise & Gaussian noise with $\sigma \in [\frac{5}{255}; \frac{14}{255}]$\\
\end{tabular}
\caption{List of implemented attack methods.}
%\vspace{-3em}
\label{table:attacks}
\end{table}

\begin{table}[h]
\centering
\begin{tabular}{lc}
Optimization target & Formula \\ 
\midrule
FTDA default & $\left\| C(x) - C(x') \right\|_{2}$ \\
Added-noises & $\left\| C(x') - C(x) - (x' - x) \right\|_{2}$  \\
Reconstruction & $\left\| C(x') - x' \right\|_{2}$ \\
FTDA MS-SSIM & MS-SSIM$\left(C(x), C(x')\right)$ \\
\makecell[l]{Reconstruction\\MS-SSIM} & MS-SSIM$\left(x', C(x')\right)$ \\
BPP increase & $1 - bpp\left(C(x')\right)$ \\
\midrule
\makecell[l]{Y-modification\\of all targets} &  Take only Y in YCbCr \\ 
\end{tabular}
\caption{List of available optimization targets, where $x$ --- original image, $x'$ --- adversarial image, $C()$ --- image reconstruction after compression}
\label{table:losses}
%\vspace{-3em}
% \end{floatrow}
\end{table}
\fi

\begin{figure*}[tb!]
\centering
\includegraphics[width=.8\textwidth]{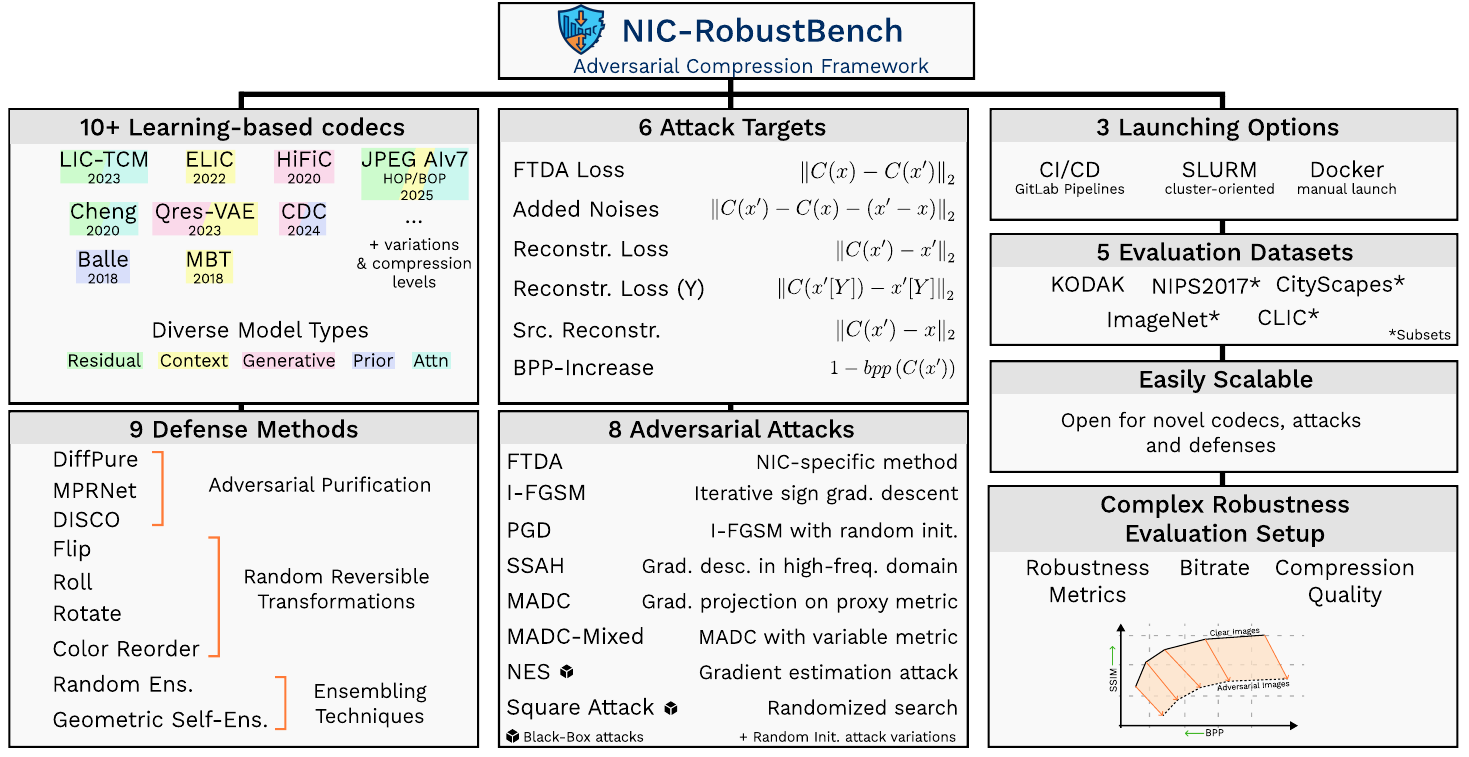}
% \caption{(a) Process of adversarial attack on NIC with possible attack goals. (b) Overall scheme of our framework.}
\caption{Summary of the contents of our framework.}
\label{fig:aa_and_scheme}
%\vspace*{-0.4cm}
\end{figure*}

\textbf{Adversarial defenses for NIC.} This work primarily focuses on transformation-based defenses, as they are universal, model-agnostic, and do not require NIC retraining, which can be prohibitively expensive for large-scale codecs. This class of defensive techniques includes a broad range of methods that augment the compression pipeline with two transformations: a preprocessing transformation $T: X \rightarrow X$ applied before compression and a postprocessing operation $T^{-1}: X \rightarrow X$ applied after decompression. The resulting defended codec is defined as $g(\cdot) = (T^{-1} \circ C \circ T)(\cdot)$, where $C$ denotes the original NIC model. An adaptive adversarial attack against the defended NIC is then formulated as
\begin{equation}
A(x) = \underset{x' : \rho(x', x) \le \varepsilon}{\arg\max} L(x, x', g(x), g(x')),
\end{equation}
where $x \in X$ is the clean input image, $x' \in X$ is the adversarially perturbed image, $\rho(\cdot,\cdot)$ denotes the perturbation constraint, and $L(\cdot)$ is the attack objective. When the transformations satisfy $T(X)=T^{-1}(X)=X$, the formulation reduces to an adversarial attack on the undefended NIC. In addition to reversible transformations (e.g., flip), we also consider neural-network-based purification defenses (e.g., DiffPure~\cite{nie2022diffusion}) that do not include a postprocessing step, i.e., $T^{-1}(X)=X$.

Other computer vision domains often consider adversarial training methods to enhance model robustness during the training stage. While adversarial training falls outside the scope of this defense formulation, adversarially trained codecs can be seamlessly added to NIC-RobustBench as independent NIC models alongside its default variants~\footnote{To this date, we have not found any publicly available NIC checkpoints trained with adversarial training techniques}.

\section{NIC-RobustBench Framework}

NIC-RobustBench is a research-oriented framework designed for systematic and comprehensive evaluation of NIC models using standardized pipelines. It provides a flexible and easily scalable test ground for further research at the intersection of adversarial robustness and learned image compression fields. Core use cases of our library include: 

% \textbf{General codec compression efficiency evaluation} in terms of image quality, compression ratio, model speed, number of parameters, and downstream CV tasks performance.

% \textbf{Attack simulation and NIC Robustness benchmarking} in a broad spectrum of adversarial scenarios with attacks using various optimization objectives that impact the compression-decompression pipeline differently.

% \textbf{Development and testing of defensive mechanisms} designed to alleviate the impact of attacks on codec performance.
% старый текст
\begin{itemize}
    \item \textbf{General codec compression efficiency evaluation} in terms of image quality, compression ratio, model speed, number of parameters, and downstream CV tasks performance.
    \item \textbf{Attack simulation and NIC Robustness benchmarking} in a broad spectrum of adversarial scenarios with attacks using various optimization objectives that impact the compression-decompression pipeline differently.
    %\item \textbf{Downstream performance evaluation} in standard CV tasks (classification, detection, depth estimation) after compression-decompression with NIC models, both in default and adversarial scenarios.
    \item \textbf{Development and testing of defensive mechanisms} designed to alleviate the impact of attacks on codec performance.
\end{itemize}

Fig. \ref{fig:aa_and_scheme} summarizes the contents of our framework. It includes the list of implemented NIC types, adversarial attacks, optimization targets for them, and adversarial defenses to counter the attacks. Our framework builds upon and extends prior research on the robustness of neural image compression methods, implementing a record number of NIC models, NIC-specific attacks, and defenses compared to existing open-source libraries (Table \ref{table:libraries}).

\begin{table}[tb]
\centering

\caption{List of popular NIC libraries and NIC robustness evaluation methods}
\resizebox{\linewidth}{!}{ 
\begin{tabular}{lccc}
Paper & \# NIC types\textbackslash variants & \# attacks & \# defenses\\ 
\midrule
\multicolumn{4}{c}{NIC robustness studies} \\
\cite{malic} & 2\textbackslash42 & 2 & 1\\
\cite{chen2023toward} & 8\textbackslash42 & 3 & 3 \\
\midrule
\multicolumn{4}{c}{Libraries}\\
\makecell[l]{CompressAI \\\cite{begaint2020compressai}} & 3\textbackslash36 & 0 & 0\\
TFC \cite{tfc_github}& 5\textbackslash- & 0 & 0 \\
\makecell[l]{NeuralCompression \\\cite{muckley2021neuralcompression}} & 5\textbackslash- & 0 & 0 \\
OpenDIC \cite{gao2024opendic} & 7\textbackslash-  & 0 & 0 \\
\textbf{NIC-RobustBench (Proposed)} & \textbf{10\textbackslash 47} & \makecell[c]{\textbf{8} \\\textbf{($\times$ 6 objectives)}} & \textbf{9}\\
\end{tabular}

}
\label{table:libraries}
%\vspace{-3em}
\end{table}

\begin{figure*}[t!]
\centering
\includegraphics[width=.95\textwidth]{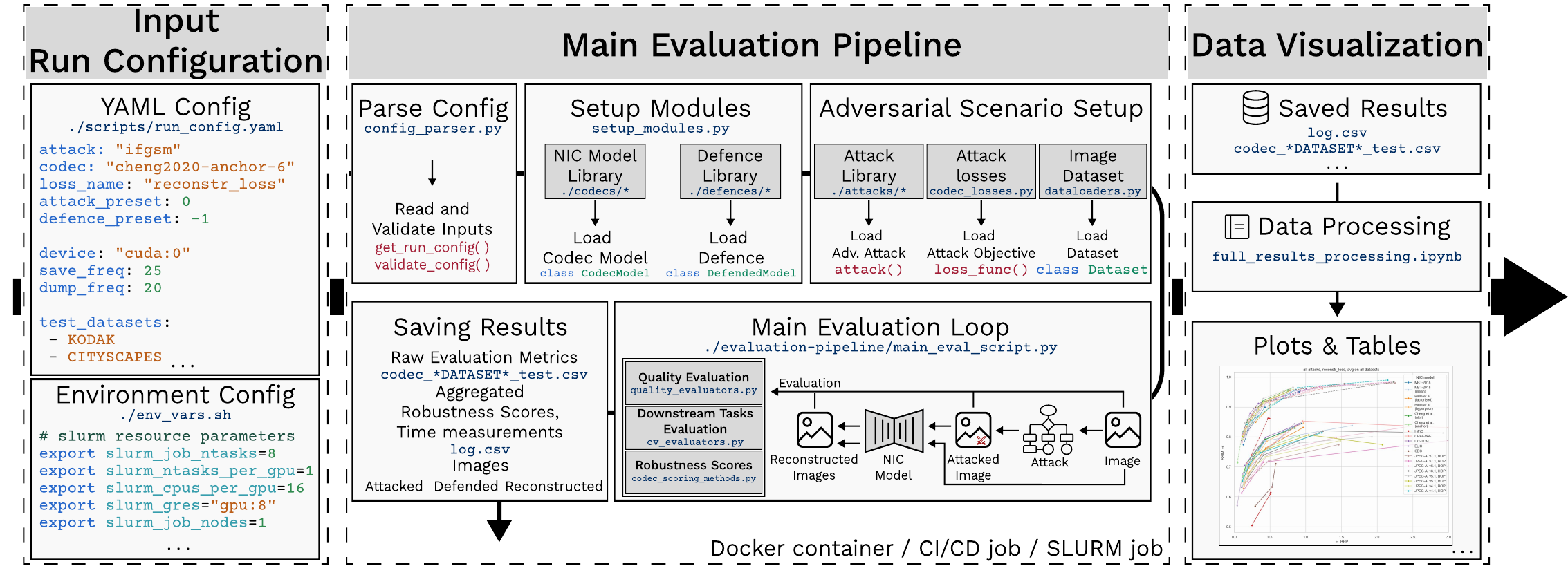}
\caption{Overview of NIC-RobustBench modular framework structure and NIC evaluation pipeline.}
\label{fig:scheme_code}
%\vspace*{-0.4cm}
\end{figure*}

\paragraph{Framework architecture.}
Fig. \ref{fig:scheme_code} illustrates the modular design and stepwise evaluation pipeline of the NIC-RobustBench framework. Its modular architecture simplifies code structure and allows for effortless scaling by implementing NIC models, datasets, adversarial attacks, and defenses as standardized classes and functions with unified methods and input arguments. Evaluation pipeline combines these modules following the input configuration and sets the adversarial scenario: for each image, a corresponding adversarial example is generated; both clean and perturbed images pass through compression–decompression and then through evaluators that record image quality, rate–distortion, timing, and downstream task performance. By leveraging Docker-based containerization and simple YAML-based setup configuration, NIC-RobustBench aims for experiment reproducibility. The framework also provides visualization utilities for quick result summaries.

% \paragraph{Framework architecture.}
% Fig. \ref{fig:scheme_code} demonstrates the modular design of the NIC-RobustBench framework and its step-by-step evaluation pipeline. The modular architecture of our library simplifies code structure and allows for effortless scaling by implementing NIC models, datasets, adversarial attacks, and defenses as standardized classes and functions with unified methods and input arguments. 

% The main robustness evaluation pipeline consists of several stages. First, it parses the input configuration, loads a selected NIC model, defensive algorithm, and dataset, and sets up the adversarial scenario. For each dataset image, a corresponding adversarial example is generated, and both images are passed through the compression-decompression process. Next, images before and after NIC pass through the evaluators that track image quality, compression efficiency, time, and downstream performance of different computer vision models. Finally, raw metric values are aggregated into evaluation scores, and all results are saved with the corresponding run metadata. 

% All stages of the pipeline are easily configurable and customizable, as core run parameters can be specified via a single YAML configuration file. By leveraging Docker-based containerization and simple YAML-based setup configuration, NIC-RobustBench aims for experiment reproducibility. Additionally, NIC-RobustBench includes tools for results visualisation, which allows researchers to quickly obtain a visual summary of the experiments.

\section{NIC Robustness Benchmark}

\textbf{NIC models.} Our framework includes most of the available open-source NIC models, including LIC-TCM \cite{liu2023learned}, ELIC \cite{he2022elic}, HiFIC \cite{mentzer2020high}, JPEG AI \cite{jpeg_ai_standard}, Cheng et al. \cite{cheng2020learned}, EVC \cite{wang2023evc}, QRes-VAE \cite{duan2023lossy}, Balle et al. \cite{balle2018variational}, MBD \cite{minnen2018joint}, and CDC \cite{yang2024lossy}. More details on these models can be found in Section \ref{sec:related_works}.

%The list of included adversarial attacks includes FTDA\cite{chen2023toward}, I-FGSM\cite{ifgsm}, MADC\cite{madc}, PGD\cite{pgd}, SSAH\cite{ssah}, CAdv\cite{cadv} and Korhonen et al.. 
\textbf{Adversarial attacks.} 
In this paper, we focus primarily on white-box attacks for several reasons: \textbf{(i)} there are no known to authors specialized black-box attacks that do not rely on transferability, and \textbf{(ii)} black-box methods typically require a large number of model queries to reach notable performance comparable to white-box attacks, making them computationally inefficient and impractical for large-scale benchmarking. In Appendix~\ref{app:query_based_attacks} we have adapted and evaluated two widely-known black-box attacks (\textbf{NES}~\cite{nes}, \textbf{Square Attack}~\cite{squareattack}), and found that they are ineffective for attacking NIC, producing only negligible degradation. 
%Moreover, Appendix~\ref{sec:transferability} studies transferability of adversarial examples across NIC models and bitrates that can be considered as a black-box attack.
%In this study, we focus on white-box attacks for the following reasons. Compression itself is a purification defense, and it can mitigate adversarial noise introduced by weaker black-box attacks, while white-box attacks offer more robust and effective perturbations. Also, black-box attacks are much more computationally expensive, severely limiting their range of applications. 
As summarized in Fig. \ref{fig:aa_and_scheme}, we have chosen six different white-box attacks of various types. \textbf{MADC} \cite{madc} was one of the first methods that uses gradient projection onto a proxy Full-Reference metric to preserve image quality. We employ 3 variations of MADC attack with different proxy metrics: $L_2$, $L_{\infty}$, and their combination, which we name MADC-Mixed. \textbf{I-FGSM} \cite{ifgsm} is a well-known iterative modification of the FGSM attack with simple sign gradient descent. \textbf{PGD} \cite{pgd} is similar to I-FGSM but uses random initialization. \textbf{FTDA} \cite{chen2023toward} attack is specifically designed to target neural image compression models. Additionally, we adopt the frequency-based \textbf{SSAH} \cite{ssah} attack that injects perturbations in high‑frequency domains to reduce visibility. %\textbf{cAdv} \cite{cadv} attack applies a filter in the LAB color space, shifting the color distribution of an image without introducing noise. We also included \textbf{Korhonen et al.}\cite{korhonen2022adversarial} as a baseline attack. It samples Gaussian noise in an attempt to attack the model.

\textbf{Adversarial defenses.} We selected several adversarial defenses to evaluate their efficiency against adaptive adversarial attacks on NIC. Previous works on NIC robustness discussed reversible transformations as defenses~\cite{chen2023toward}. Therefore, we employ reversible \textbf{Flip}, \textbf{Random Roll}, \textbf{Random Rotate}, and \textbf{Random Color Reorder} in addition to the following transformations. \textbf{Random Ensemble} mixes Roll, Rotate, and Color Reorder. It samples and applies 10 actions from Roll, Rotate, and Color reorder with 4, 4, and 1 weights, respectively. \textbf{Geometric Self-Ensemble} \cite{chen2023toward} generates 8 flip/rotation variants of the input image. It runs each one through the preprocess-NIC-postprocess pipeline and selects the candidate with minimal distortion in terms of mean squared error (MSE) to the original. In our paper, we additionally employed purification defenses that are applied for similar tasks. \textbf{DiffPure} \cite{nie2022diffusion} performs purification based on a diffusion model as preprocessing and does nothing in the postprocessing step. \textbf{DISCO}~\cite{disco} is an adversarial purification defense that reconstructs a clean image using per‑pixel features and a neighborhood‑conditioned module. \textbf{MPRNet}~\cite{mprnet} is a multi-stage CNN for image restoration with cross‑stage feature fusion that can also be used as a purification method.

\textbf{Datasets.} To evaluate methods, we chose five well-known datasets. The KODAK Photo CD \cite{kodakDataset} dataset consists of 24 uncompressed images, each with a resolution of $768 \times 512$. We sample 50 $512\times256$ images from CITYSCAPES \cite{cityscapesDataset}, which is a domain-specific dataset for image segmentation tasks. These datasets are commonly used in the field of image compression. Additionally, NIPS 2017: Adversarial Learning Development Set \cite{nipsDataset} contains 1000 $299\times299$ images and is designed for evaluating adversarial attacks against image classifiers. %Finally, the BSDS dataset \cite{bsdsDataset}, with 500 images of $320 \times 448$ resolution focusing on segmentation and boundary detection, enables us to evaluate the effects of NIC on segmentation model performance, further assessing implications of attacks on compression models for computer vision tasks.
Finally, in Appendix~\ref{app:additional_results}, we evaluated high-resolution datasets: ImageNet and CLIC, among others. ImageNet~\cite{deng2009imagenet} is a large-scale collection of human-labeled images that became a standard benchmark (and pretraining source) for visual recognition models. CLIC~\cite{clic2025} is a curated dataset released for the Challenge on Learned Image Compression, used to evaluate and compare learned end-to-end image compression methods under a shared benchmark and leaderboard.

\textbf{Evaluation metrics.}
We employ four different Full-Reference image quality metrics to numerically assess the effects of adversarial attacks on images before and after the reconstruction: PSNR, MSE, MS-SSIM \cite{ms_ssim}, and VMAF \cite{li2018vmaf}. PSNR, MSE, and MS-SSIM are traditional image-similarity measures. MS-SSIM \cite{ms_ssim} provides a scale-independent quality estimate, and VMAF implements a learning-based approach that aligns well with human perception \cite{NEURIPS2022_59ac9f01}. VMAF was designed to estimate the quality of distorted videos, but it can also be applied to images, interpreting them as single-frame videos.

Following the methodology of \cite{chen2023toward}, we measure $\Delta_{score}$~, which quantifies how adversarial distortion changes after image reconstruction by a NIC model:
\begin{equation}
 \Delta_{score}=FR(x_i, x_i') - FR(C(x_i), C(x_i')),
\end{equation}
where $x$ is an original image, $x'$ is a corresponding adversarial example, $FR(x,y)$ is one of the aforementioned image quality metrics, and $C(x)$ is an evaluated NIC model (entire encoding-decoding procedure). For metrics where larger $FR$ values indicate better visual quality (e.g., for PSNR, MS-SSIM, and VMAF), then positive $\Delta_{score}$ indicates that the input perturbation was amplified by the model, whereas negative values indicate that it was suppressed.

Additionally, we introduce $\delta_{score}$ --- a different measure that captures the difference in reconstruction quality between the clear image and corresponding adversarial example:
\begin{equation}
 \delta_{score}= FR(x, C(x)) - FR(x', C(x')),
\end{equation}
High $\delta_{score}$ indicates that the adversarial example introduced significant distortions to the decompressed image after the NIC. Conversely, near-zero values suggest that compression degrades clean and adversarial inputs similarly. Unlike $\Delta_{score}$, $\delta_{score}$ measures only post‑compression quality loss and ignores attack visibility/imperceptibility. Hence, it is well-suited for robustness evaluation in applications with minimum quality requirements and for assessing attacks without perturbation‑strength constraints.

% To measure transferability of adversarial attacks to other codecs, we measure a modified version of $\Delta_{score}$:
% \begin{equation}
% \label{eq:delta_transf}
%  \hat{\Delta}_{score}=FR(C(x_i), C(x_i')) - FR(C'(x_i), C'(x_i')),
% \end{equation}
% where $C(x)$ is the decoded image by the target codec used to construct an adversarial image, and $C'(x)$ is the attacked codec to which we test the attack's transferability. This metric offers a more precise evaluation for different pairs of models when comparing NIC with different robustness. 

\section{Evaluation Results}
% \begin{figure*}[t!]
% \centering
% \includegraphics[width=.99\textwidth]{iclr2026/images/results/lineplots_losses.pdf}
% \caption{RD-curves representing NIC performance on clear~(dashed lines) and attacked~(solid) data. Attacks target image quality on the left subfigure, and bitrate on the right.}
% \label{fig:lineplot_2_losses}
% %\vspace*{-0.4cm}
% \end{figure*}
\begin{figure*}[t!]
\centering
\includegraphics[width=.95\textwidth]{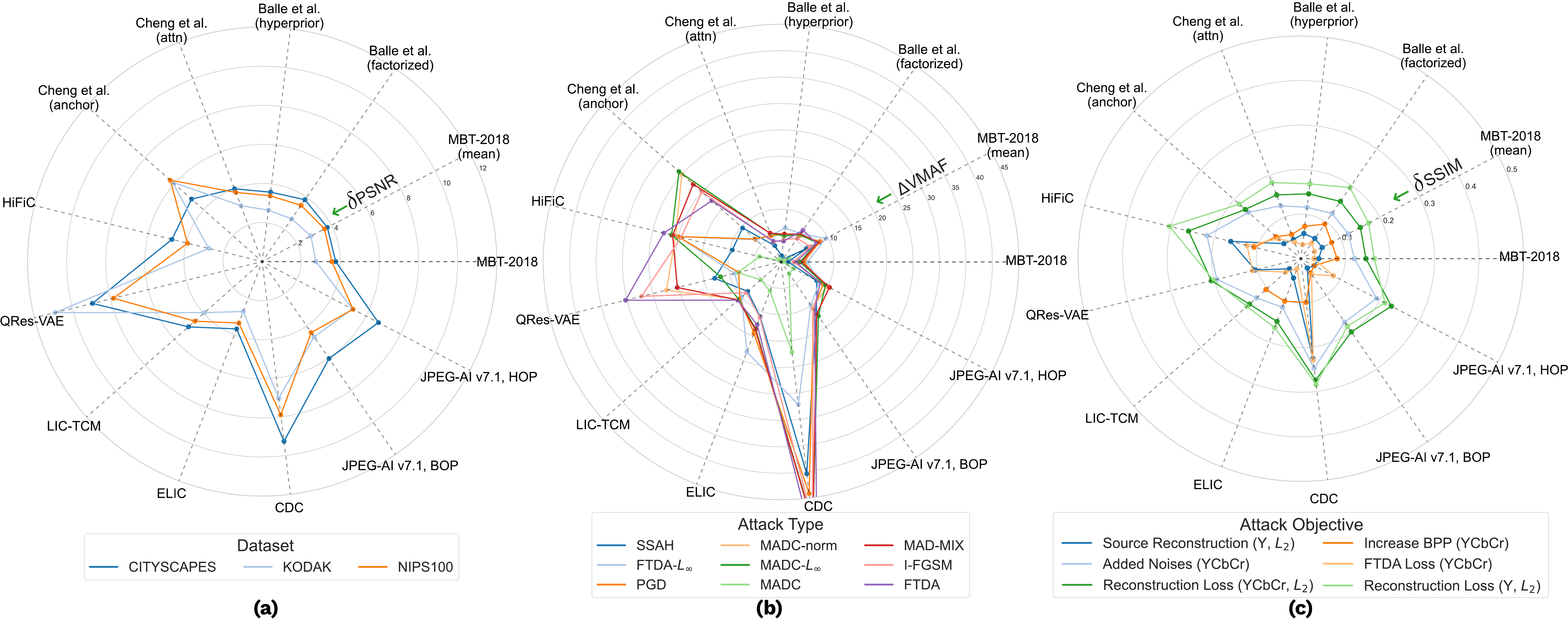}
\caption{Robustness evaluation of NIC models measured across (a) different datasets, (b) different attacks, and (c) different objectives.}
\label{fig:roseplots}
%\vspace*{-0.4cm}
\end{figure*}

\subsection{NIC performance evaluation}

%Figure \ref{} presents the Rate-Distortion curves~(SSIM/BPP) both in clear and adversarial scenarios. Dashed RD-curves represent the performance of different NICs on clear images.

% \textbf{Generative types of codec lead without attacks.} In our evaluations, codecs that use generative approaches, like CDC, and codecs with hyperprior architectures, like LIC-TCM, demonstrate the best quality-bitrate ratio at lower bitrates. At higher bitrates, the attention architectures, notably JPEG AI and Cheng et al., produce better images than the others, tied with generative type. 

%Fig. \ref{fig:roseplots} (a-c) demonstrates the impact of adversarial attacks on NIC models and their reconstruction quality in various data slices. As Fig. \ref{fig:roseplots} (a) suggests, evaluation scores are mostly consistent across multiple image sets of varying resolutions, aspect ratios, and contents.

\begin{figure*}[t!]
\centering
\includegraphics[width=.95\textwidth]{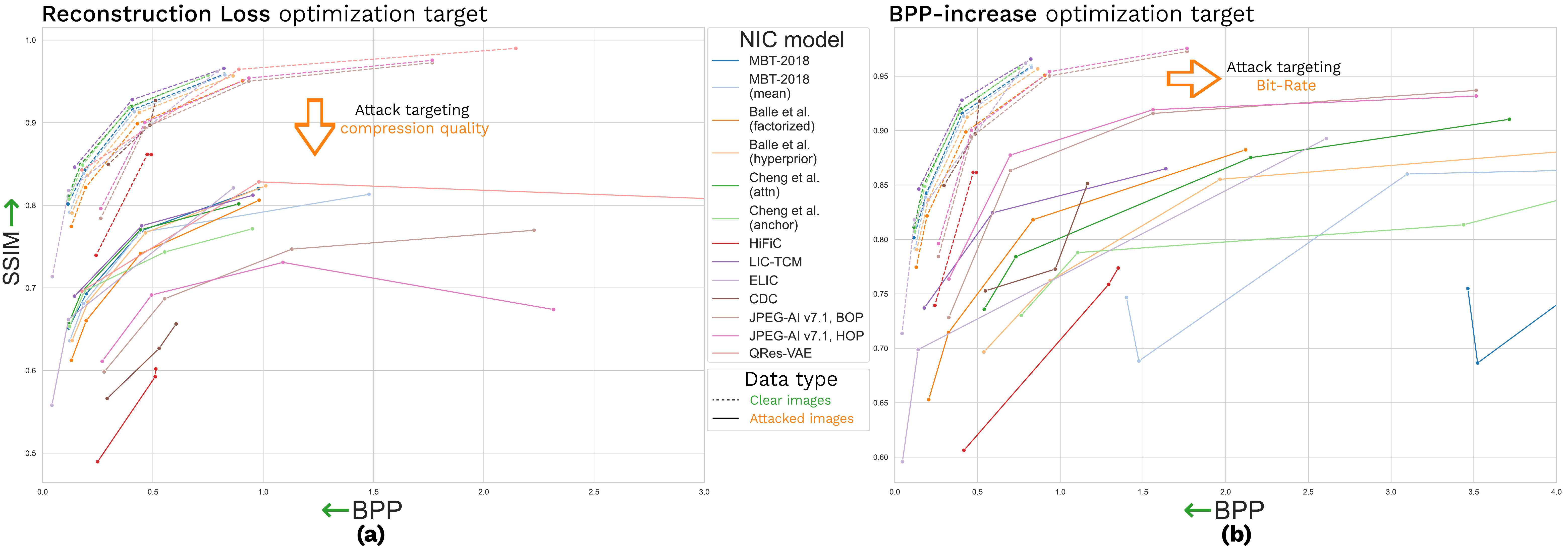}
\caption{RD-curves representing NIC performance on clear~(dashed lines) and attacked~(solid) data. Attacks target image quality on the left subfigure, and bitrate on the right.}
\label{fig:lineplot_2_losses}
%\vspace*{-0.4cm}
\end{figure*}

The main results of NIC performance evaluation are presented in Fig. \ref{fig:roseplots} and \ref{fig:lineplot_2_losses}. Fig. \ref{fig:roseplots} illustrates the impact of adversarial attacks on the reconstruction quality of neural compression models across different data slices. Fig. \ref{fig:lineplot_2_losses} compares the performance of neural codecs on clean and adversarially perturbed data under different target objectives used in the attacks. Based on these results, the following insights can be drawn.

\begin{figure}[tb]
    \centering
    \includegraphics[width=0.95\linewidth]{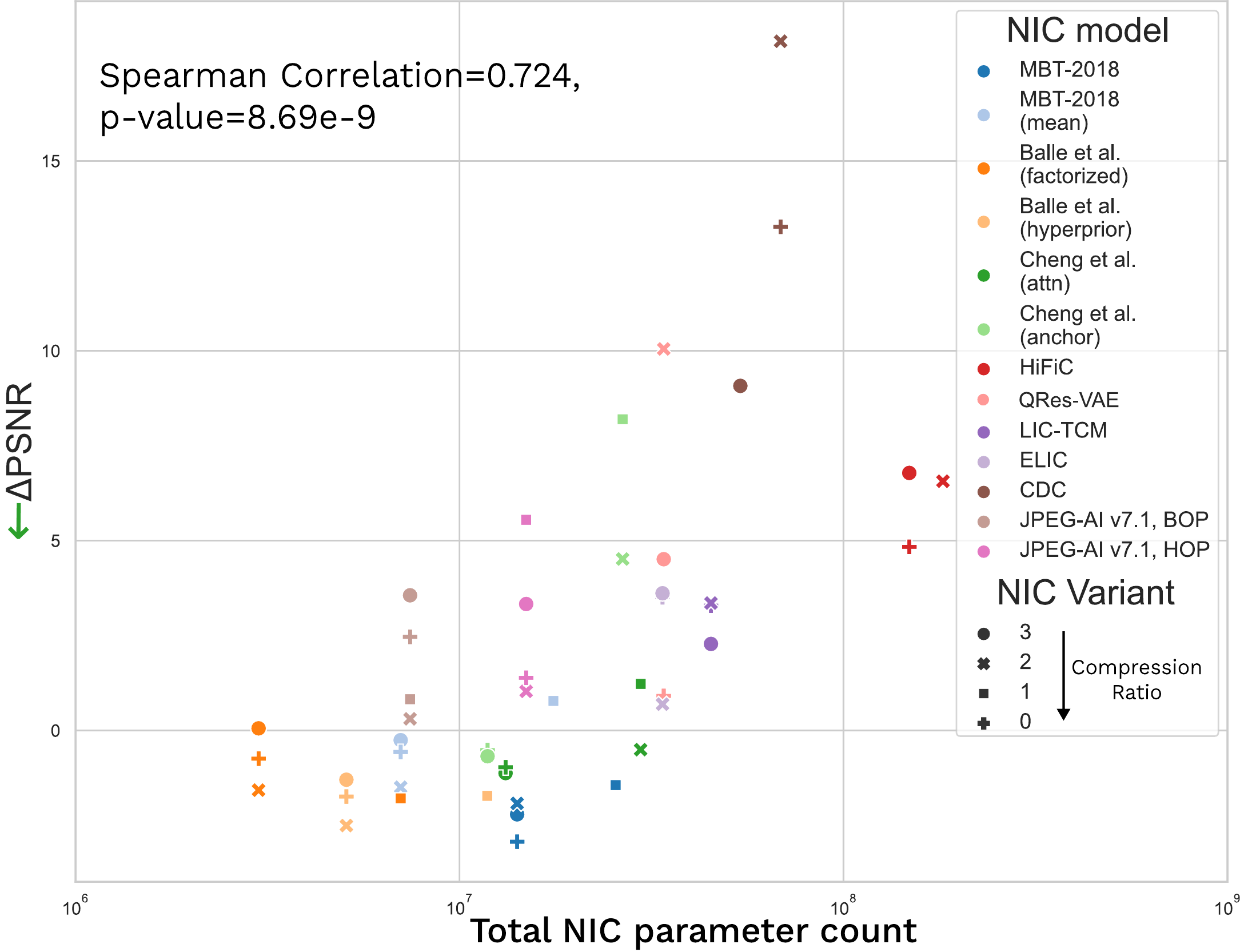}
    \caption{Relationship between NIC size, compression ratio, and robustness.}
    \label{fig:corr_plot}
    %\vspace{-5pt}
\end{figure}

\textbf{Generative codecs are among the most vulnerable to attacks.} 
Codecs with generative priors in their design, namely CDC, HiFiC, and QRes-VAE, exhibit high $\Delta$ and $\delta$ scores (Fig. \ref{fig:roseplots}), suggesting stronger adversarial effects. HiFiC uses GAN in its architecture to train adversarially against the discriminator, CDC employs a diffusion model, and QRes-VAE uses a Variational AutoEncoder. However, we also note that this effect might be attributed to the larger size of these models compared to other codecs, as described below. We attribute this vulnerability to the way generative codecs rely on a latent representation of the entire image: even small perturbations can shift the latents globally, affecting the reconstructed image as a whole. In contrast, discriminative codecs operate primarily through local, pixel-wise prediction mechanisms. Because they do not depend on a global latent space, perturbations might have a more localized effect and do not propagate through the entire representation, which makes these codecs inherently more robust. Fig.~\ref{fig:corr_plot} shows that QRes-VAE (generative) and ELIC (discriminative) models have a comparable number of parameters, confirming our finding. 

\textbf{Across NIC model families, increased model capacity correlates strongly with reduced adversarial robustness}. Throughout our study, we found a significant (Spearman Corr. $0.724$, $p<10^{-8}$) positive correlation between model size and average efficiency of the attacks, as presented in Fig. \ref{fig:corr_plot}. While the scale of the effect varies from model to model, a larger parameter count appears to create more pathways within the model that attacks could potentially exploit. Fig. \ref{fig:roseplots} also confirms that smaller Balle et al. models and MBT-2018 are among the most resilient, while larger HiFiC and CDC are among the most vulnerable. Larger models generally achieve better rate–distortion performance on clean images by allocating more capacity and bit budget to fine, high-frequency details. However, this also means that the models may have a greater number of parameters than required for stable performance; therefore, they exhibit lower generalization and robustness. In contrast, smaller or low‑BPP variants act like low‑pass filters, prioritizing coarse structure and partially suppressing adversarial perturbations, thus exhibiting higher robustness.

\textbf{Within a NIC model family, those with higher compression rates are the most robust}. These models (with lower BPPs, see variants 0 and 1 in Fig. \ref{fig:corr_plot}) typically have fewer parameters and are thus less vulnerable. Moreover, smaller models targeting stronger compression often introduce image blur by prioritizing lower frequencies over higher ones, which can ``erase'' adversarial perturbations more effectively during compression. Therefore, this effect may be attributed not only to the reduced number of parameters but also to the frequency characteristics of the compression process.

\textbf{Robustness scores are mostly consistent across images with different resolutions, aspect ratios, and content}. The evaluation of NIC models on three distinct datasets, presented in Fig. \ref{fig:roseplots}~(a), revealed a nearly identical robustness profile, with no substantial differences observed across datasets. Experiments in the Appendix (Sec. ~\ref{app:additional_results}) compare the results on those datasets with high-resolution large-scale datasets (ImageNet and CLIC) and demonstrate the generalization of our findings across the chosen datasets.

%Model variants trained for higher compression ratios (i.e., lower BPPs, variants 0,1 in Fig. \ref{fig:corr_plot}), that usually have fewer parameters, are generally more robust. Smaller models that aim for strong compression also typically introduce blur into the image, as they prioritise lower frequencies over the higher ones, which may “erase” adversarial perturbations more efficiently during the compression.

\textbf{Best-performing NICs in the non-adversarial scenario vary across different bitrates.} Alongside the NIC robustness comparison, our framework can also be used to directly compare NIC models on non-attacked images, like other NIC libraries. Dashed rate-distortion (RD) curves in Fig. \ref{fig:lineplot_2_losses} represent the performance of different NICs on source images. In the lowest bitrate range ($<0.2$ BPP), the context-based ELIC model leads, as it reaches the lowest bitrate among the tested models. In the most competitive lower-to-mid bitrate range ($0.2-0.8$ BPP), NIC models that incorporate attention mechanisms, namely LIC-TCM and Cheng2020-attn, show the best efficiency in terms of SSIM. LIC-TCM combines transformer blocks with a CNN backbone, and Cheng2020-attn enhances a residual network with Self-Attention. For the highest bitrate scenarios, the VAE-based QRes-VAE model shows the best quality.

%\textbf{X codecs are the fastest.}

\subsection{Attack efficiency analysis}
\textbf{Distinct attack objectives have vastly different impacts on the target NIC model.} Fig. \ref{fig:lineplot_2_losses} compares attacks targeting two ``orthogonal'' measures of codec performance: rate (right subfigure) and distortion (left). Both types cause substantial NIC performance drop; however, the direction of the impact differs. Reconstruction attack objective severely impairs the quality of reconstructed images compared to their uncompressed variants, shifting RD-curves downwards. BPP-increase objective, on the other hand, targets compressed file size, and shifts RD-curves to the right, towards higher BPP. At the same time, models that are robust against attacks based on one objective function may become vulnerable when a different objective is used. For example, the MBT-2018 family demonstrated strong robustness under the Reconstruction attack objective but proved to be the most vulnerable under the Increase-BPP target.

\textbf{Attacks directly targeting image reconstruction have the strongest impact on NIC performance in terms of image quality.} As illustrated in Fig. \ref{fig:roseplots} (c), attack objectives that maximize the distance between attacked images before and after compression (``Reconstruction Loss'' in Fig. \ref{fig:roseplots}, i.e. $||C(x’)-x’||_2$) result in the highest loss of SSIM. Other targets that include a clear image $x$ or its decompressed variant $C(x)$ to the objective show a weaker impact on quality. Moreover, the most effective variant is reconstruction loss applied only to the luminance channel, which enabled stronger disruption of the reconstructed image structure. 

\textbf{Attack efficiency varies widely across codecs.} As Fig. \ref{fig:roseplots} (b) demonstrates, FTDA and MADC-$L_{\infty}$ are among the strongest in our study, showing the highest $\Delta$ scores on most codecs. However, two exceptions are notable: on the Cheng2020 (anchor) model, FTDA weakens significantly, while MADC-based attacks and I-FGSM perform well. In contrast, on QRes-VAE, FTDA significantly outperforms MADC-$L_{\infty}$. This effect might be caused by gradient quantization: MADC-$L_{\infty}$ attack uses a $sign$ operation on the gradients, while FTDA uses an unmodified gradient in the optimizer. This gradient quantization step might speed up the attack convergence for some models (e.g., Cheng2020 (anchor)), while significantly impairing the accuracy of the optimization steps for others (e.g., QRes-VAE). This emphasizes that NIC models should be tested against various attack designs to ensure better robustness.

\subsection{Defense evaluation}

\begin{figure}[tb]
    \centering
    \includegraphics[width=0.95\linewidth]{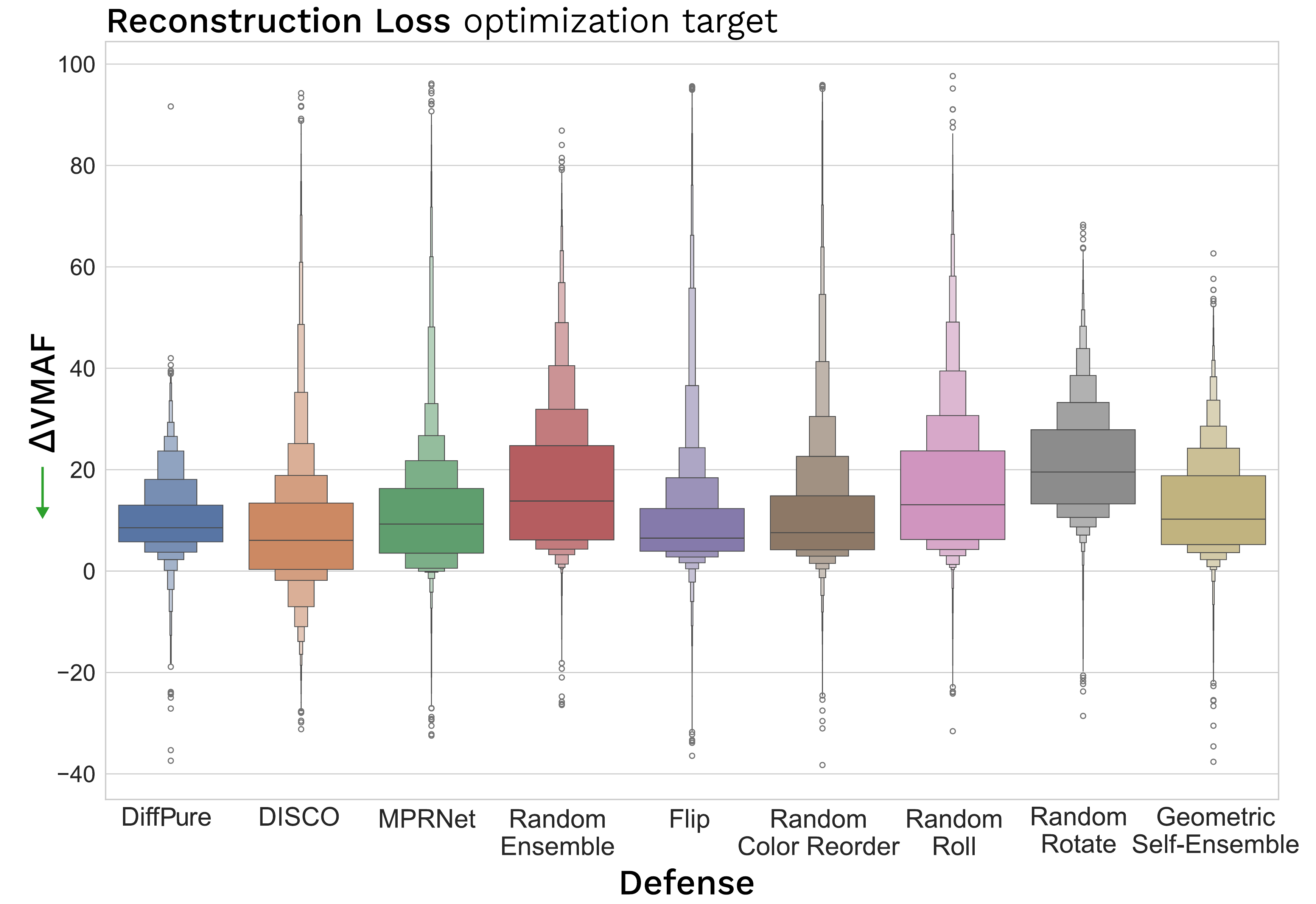}
    \caption{Performance of different defensive algorithms.}
    \label{fig:defenses}
\end{figure}
% \begin{figure*}[t!]
% \centering
% \includegraphics[width=.99\textwidth]{iclr2026/images/results/defences_boxen.pdf}
% \caption{Performance of different defensive algorithms}
% \label{fig:roseplots}
% %\vspace*{-0.4cm}
% \end{figure*}
\textbf{DiffPure and DISCO are among the strongest adversarial defenses.}
As Fig. \ref{fig:defenses} shows, DISCO defense results in better performance on average, while for DiffPure, $\Delta$VMAF scores are more concentrated compared to other defenses and lack outliers at the top, hinting that it is more stabilizing and predictable for NIC models. Both of these defensive models are designed to remove off-manifold, high-frequency adversarial noise, pulling inputs back toward the data distribution, which might explain their efficiency.

\textbf{Geometric transformations without intrinsic interpolation work better}. As shown in Fig. \ref{fig:defenses}, a simple flip is effective: it is a lossless transform aligned with the model’s learned invariance from training, so perturbations don’t transfer while semantics remain intact. By contrast, Random Rotate and Random Roll introduce interpolation/edge artifacts and a distribution shift that the model is not equivariant to, hurting clean accuracy and capping robust gains.

\textbf{For some NICs, adversarial defenses show limited RD-performance gains}. As demonstrated in Fig. \ref{fig:lineplots_defenses} in the Appendix, NIC models without defenses in many cases outperform defended variants in Rate-Distortion efficiency even on adversarial data. Irreversible learning-based defenses (e.g., MPRNet, DISCO), while erasing the adversarial noise, may introduce their own artifacts into the image, which may further propagate through the compression-decompression process. This leads to reduced image quality compared to the original non-processed image, and in some cases can also inflate the bitrate. This finding highlights the importance of further development of compression-specific defensive techniques, as traditional methods designed mostly for classification and other discriminative vision tasks may show limited or even negative effects in a more specific and fine-grained domain, such as learned image compression.

\subsection{Additional results}

We provide more detailed results in the Appendix, including attack transferability evaluation between different NIC models (Sec. \ref{sec:transferability}), computational performance evaluations (Sec. \ref{sec:speed}), statistical tests (Sec. \ref{sec:wilcoxon}), attacks on downstream computer vision tasks (Sec. \ref{sec:cv_tasks}), query-based black-box attacks on NIC (Sec.~\ref{app:query_based_attacks}), and other experiments.

\section{Conclusion}

This paper proposes NIC-RobustBench, a novel open-source benchmark and library to evaluate NIC robustness. It is the first large-scale framework that implements a comprehensive set of adversarial attacks, objectives, and defenses for the learned image compression task, covering the largest collection of NIC models among existing open-source libraries known to the authors. Beyond providing an efficient, simple, and scalable framework, NIC-RobustBench enables the analysis of vulnerabilities across architectures, attack regimes, and compression rates. With this framework, we conduct extensive experiments across various datasets and identify several insights on NIC robustness, including the vulnerability of large and generative codecs, stability of low-bitrate models, and the limited effectiveness of generic vision defenses in compression pipelines --- highlighting the importance of robustness validation in the image-compression task.

%This paper proposes NIC-RobustBench, a novel open-source library and benchmark that researchers can use for evaluating NIC robustness. NIC-RobustBench is the first large-scale framework that implements a comprehensive collection of adversarial attacks and defenses for the learned image compression task. It includes the largest number of NIC models among the open-source libraries known to the authors. NIC-RobustBench provides an efficient, simple, and scalable framework to evaluate the performance, robustness, and compression quality of NIC models. Using NIC-RobustBench, we reveal systematic degradation of compression quality and efficiency in most modern learning-based codecs under adversarial attacks, emphasizing the importance of robustness validation in the image-compression task.

\section*{Impact Statement}
This paper presents work whose goal is to advance the field of Machine Learning. There are potential societal consequences, none of which we feel must be highlighted here.

\bibliography{example_paper}

@String(CVPR= {IEEE Conf. Comput. Vis. Pattern Recog.})

@String(ICCV= {Int. Conf. Comput. Vis.})

@String(ECCV= {Eur. Conf. Comput. Vis.})

@String(NIPS= {Adv. Neural Inform. Process. Syst.})

@String(ICLR = {Int. Conf. Learn. Represent.})

@String(AAAI = {AAAI})

@String(CVPR  = {CVPR})

@String(ICCV  = {ICCV})

@String(ECCV  = {ECCV})

@String(NIPS  = {NeurIPS})

@String(ICLR  = {ICLR})

@inproceedings{duan2023lossy,
  title={Lossy image compression with quantized hierarchical vaes},
  author={Duan, Zhihao and Lu, Ming and Ma, Zhan and Zhu, Fengqing},
  booktitle={Proceedings of the WACV},
  pages={198--207},
  year={2023}
}

@article{malic,
  title={Manipulation attacks on learned image compression},
  author={Liu, Kang and Wu, Di and Wu, Yangyu and Wang, Yiru and Feng, Dan and Tan, Benjamin and Garg, Siddharth},
  journal={IEEE Transactions on Artificial Intelligence},
  volume={5},
  number={6},
  pages={3083--3097},
  year={2023},
  publisher={IEEE}
}

@inproceedings{ifgsm,
  title={Adversarial examples in the physical world},
  author={Kurakin, Alexey and Ian J. Goodfellow and Samy Bengio},
  booktitle={ICLR Workshop},
  year={2017}
}

@inproceedings{madc,
  title={Maximum differentiation (MAD) competition: A methodology for comparing computational models of perceptual quantities},
  author={Wang, Zhou and Eero P. Simoncelli},
  booktitle={Journal of Vision 8.12},
  pages={8--8},
  year={2008}
}

@inproceedings{cw,
    title={Towards evaluating the robustness of neural networks},
    author={Carlini, Nicholas and Wagner, David},
    booktitle={2017 IEEE Symposium on Security and Privacy},
    year={2017},
}

@inproceedings{pgd,
    title={Towards Deep Learning Models Resistant to Adversarial Attacks},
    author={Aleksander Madry and Aleksandar Makelov and Ludwig Schmidt and Dimitris Tsipras and Adrian Vladu},
    booktitle={ICLR},
    year={2018},
}

@inproceedings{ssah,
  title={Frequency-driven imperceptible adversarial attack on semantic similarity},
  author={Luo, Cheng and Lin, Qinliang and Xie, Weicheng and Wu, Bizhu and Xie, Jinheng and Shen, Linlin},
  booktitle={Proceedings of the CVPR},
  pages={15315--15324},
  year={2022}
}

@article{cadv,
  title={Unrestricted adversarial examples via semantic manipulation},
  author={Bhattad, Anand and et al},
  journal={arXiv preprint arXiv:1904.06347},
  year={2019}
}

@inproceedings{liu2023learned,
  title={Learned image compression with mixed transformer-cnn architectures},
  author={Liu, Jinming and Sun, Heming and Katto, Jiro},
  booktitle={Proceedings of the CVPR},
  pages={14388--14397},
  year={2023}
}

@inproceedings{wang2023evc,
    title={Evc: Towards real-time neural image compression with mask decay},
    author={Wang, Guo-Hua and Li, Jiahao and Li, Bin and Lu, Yan},
    booktitle={ICLR},
    year={2023},
}

@article{yang2024lossy,
  title={Lossy image compression with conditional diffusion models},
  author={Yang, Ruihan and Mandt, Stephan},
  journal={NeurIPS},
  volume={36},
  year={2024}
}

@article{mentzer2020high,
  title={High-fidelity generative image compression},
  author={Mentzer, Fabian and Toderici, George D and Tschannen, Michael and Agustsson, Eirikur},
  journal={NeurIPS},
  volume={33},
  year={2020}
}

@inproceedings{he2022elic,
  title={Elic: Efficient learned image compression with unevenly grouped space-channel contextual adaptive coding},
  author={He, Dailan and Yang, Ziming and Peng, Weikun and Ma, Rui and Qin, Hongwei and Wang, Yan},
  booktitle={Proceedings of the CVPR},
  pages={5718--5727},
  year={2022}
}

@article{begaint2020compressai,
  title={Compressai: a pytorch library and evaluation platform for end-to-end compression research},
  author={B{\'e}gaint, Jean and Racap{\'e}, Fabien and Feltman, Simon and Pushparaja, Akshay},
  journal={arXiv preprint arXiv:2011.03029},
  year={2020}
}

@inproceedings{cheng2020learned,
  title={Learned image compression with discretized gaussian mixture likelihoods and attention modules},
  author={Cheng, Zhengxue and Sun, Heming and Takeuchi, Masaru and Katto, Jiro},
  booktitle={Proceedings of CVPR},
  pages={7939--7948},
  year={2020}
}

@article{minnen2018joint,
  title={Joint autoregressive and hierarchical priors for learned image compression},
  author={Minnen, David and Ball{\'e}, Johannes and Toderici, George D},
  journal={NeurIPS},
  volume={31},
  year={2018}
}

@inproceedings{zou2022devil,
  title={The devil is in the details: Window-based attention for image compression},
  author={Zou, Renjie and Song, Chunfeng and Zhang, Zhaoxiang},
  booktitle={CVPR},
  pages={17492--17501},
  year={2022}
}

@article{balle2016end,
  title={End-to-end optimized image compression},
  author={Ball{\'e}, Johannes and Laparra, Valero and Simoncelli, Eero P},
  journal={arXiv preprint arXiv:1611.01704},
  year={2016}
}

@article{agustsson2017soft,
  title={Soft-to-hard vector quantization for end-to-end learning compressible representations},
  author={Agustsson, Eirikur and Mentzer, Fabian and Tschannen, Michael and Cavigelli, Lukas and Timofte, Radu and Benini, Luca and Gool, Luc V},
  journal={NeurIPS},
  volume={30},
  year={2017}
}

@inproceedings{balle2018variational,
  title={Variational image compression with a scale hyperprior},
  author={Ball{\'e}, Johannes and Minnen, David and Singh, Saurabh and Hwang, Sung Jin and Johnston, Nick},
  booktitle={ICLR},
  year={2018}
}

@inproceedings{cityscapesDataset,
title={The Cityscapes Dataset for Semantic Urban Scene Understanding},
author={Cordts, Marius and Omran, Mohamed and Ramos, Sebastian and Rehfeld, Timo and Enzweiler, Markus and Benenson, Rodrigo and Franke, Uwe and Roth, Stefan and Schiele, Bernt},
booktitle={CVPR},
year={2016}
}

@article{kodakDataset,
  title={Kodak lossless true color image suite (photocd pcd0992)},
  author={Kodak, Eastman},
  journal={URL http://r0k. us/graphics/kodak},
  volume={6},
  number={2},
  pages={5},
  year={1993}
}

@incollection{nipsDataset,
  title={Adversarial attacks and defences competition},
  author={Kurakin, Alexey and Goodfellow, Ian and Bengio, Samy and Dong, Yinpeng and Liao, Fangzhou and Liang, Ming and Pang, Tianyu and Zhu, Jun and Hu, Xiaolin and Xie, Cihang and others},
  booktitle={The NIPS'17 Competition: Building Intelligent Systems},
  pages={195--231},
  year={2018},
  publisher={Springer}
}

@article{nie2022diffusion,
  title={Diffusion models for adversarial purification},
  author={Nie, Weili and Guo, Brandon and Huang, Yujia and Xiao, Chaowei and Vahdat, Arash and Anandkumar, Anima},
  journal={arXiv preprint arXiv:2205.07460},
  year={2022}
}

@inproceedings{dong2018boosting,
  title={Boosting adversarial attacks with momentum},
  author={Dong, Yinpeng and Liao, Fangzhou and Pang, Tianyu and Su, Hang and Zhu, Jun and Hu, Xiaolin and Li, Jianguo},
  booktitle={CVPR},
  year={2018}
}

@article{chakraborty2021survey,
  title={A survey on adversarial attacks and defences},
  author={Chakraborty, Anirban and Alam, Manaar and Dey, Vishal and Chattopadhyay, Anupam and Mukhopadhyay, Debdeep},
  journal={CAAI Transactions on Intelligence Technology},
  volume={6},
  number={1},
  pages={25--45},
  year={2021},
  publisher={Wiley Online Library}
}

@article{nezami2021pick,
  title={Pick-object-attack: Type-specific adversarial attack for object detection},
  author={Nezami, Omid Mohamad and Chaturvedi, Akshay and Dras, Mark and Garain, Utpal},
  journal={Computer Vision and Image Understanding},
  volume={211},
  pages={103257},
  year={2021},
  publisher={Elsevier}
}

@article{iqarobustness,
title={Comparing the robustness of modern no-reference image- and video-quality metrics to adversarial attacks},
author={Antsiferova, Anastasia and Abud, Khaled and Gushchin, Aleksandr and Shumitskaya, Ekaterina and Lavrushkin, Sergey and Vatolin, Dmitriy},
journal={AAAI},
volume={38},
url={https://ojs.aaai.org/index.php/AAAI/article/view/27827},
DOI={10.1609/aaai.v38i2.27827},
number={2},
year={2024},
month={Mar.},
pages={700-708}
}

@article{nic1,
  title={Improving inference for neural image compression},
  author={Yang, Yibo and Bamler, Robert and Mandt, Stephan},
  journal={NeurIPS},
  volume={33},
  pages={573--584},
  year={2020}
}

@inproceedings{nic2,
  title={Slimmable compressive autoencoders for practical neural image compression},
  author={Yang, Fei and Herranz, Luis and Cheng, Yongmei and Mozerov, Mikhail G},
  booktitle={CVPR},
  year={2021}
}

@inproceedings{nic3,
  title={Neural image compression via attentional multi-scale back projection and frequency decomposition},
  author={Gao, Ge and You, Pei and Pan, Rong and Han, Shunyuan and Zhang, Yuanyuan and Dai, Yuchao and Lee, Hojae},
  booktitle={Proceedings of ICCV},
  pages={14677--14686},
  year={2021}
}

@article{li2018vmaf,
  title={VMAF: The journey continues},
  author={Li, Zhi and Bampis, Christos and Novak, Julie and Aaron, Anne and Swanson, Kyle and Moorthy, Anush and Cock, JD},
  journal={Netflix Technology Blog},
  volume={25},
  year={2018}
}

@inproceedings{
NEURIPS2022_59ac9f01,
author = {Antsiferova, Anastasia and Lavrushkin, Sergey and Smirnov, Maksim and Gushchin, Aleksandr and Vatolin, Dmitriy and Kulikov, Dmitriy},
booktitle = {NeurIPS},
title = {Video compression dataset and benchmark of learning-based video-quality metrics},
year = {2022}
}

@ARTICLE{jpeg_ai_standard,
  author={Ascenso, João and Alshina, Elena and Ebrahimi, Touradj},
  journal={IEEE MultiMedia}, 
  title={The JPEG AI Standard: Providing Efficient Human and Machine Visual Data Consumption}, 
  year={2023},
  volume={30},
  number={1},
  pages={100-111},
  doi={10.1109/MMUL.2023.3245919}}

@software{tfc_github,
  author = "Ballé, Jona and Hwang, Sung Jin and Agustsson, Eirikur",
  title = "{T}ensor{F}low {C}ompression: Learned Data Compression",
  url = "http://github.com/tensorflow/compression",
  version = "2.14.1",
  year = "2024",
}

@misc{muckley2021neuralcompression,
    author={Matthew Muckley and Jordan Juravsky and Daniel Severo and Mannat Singh and Quentin Duval and Karen Ullrich},
    title={NeuralCompression},
    howpublished={\url{https://github.com/facebookresearch/NeuralCompression}},
    year={2021}
}

@inproceedings{gao2024opendic,
  title={OpenDIC: An Open-Source Library and Performance Evaluation for Deep-learning-based Image Compression},
  author={Gao, Wei and Zheng, Huiming and Zhang, Chenhao and Zheng, Kaiyu and Yu, Zhuozhen and Li, Yuan and Ye, Hua and Zhang, Yongchi},
  booktitle={ACM MM},
  pages={11202--11205},
  year={2024}
}

@article{croce2020robustbench,
  title={Robustbench: a standardized adversarial robustness benchmark},
  author={Croce, Francesco and Andriushchenko, Maksym and Sehwag, Vikash and Debenedetti, Edoardo and Flammarion, Nicolas and Chiang, Mung and Mittal, Prateek and Hein, Matthias},
  journal={arXiv preprint arXiv:2010.09670},
  year={2020}
}

@article{chen2023toward,
  title={Toward Robust Neural Image Compression: Adversarial Attack and Model Finetuning},
  author={Chen, Tong and Ma, Zhan},
  journal={IEEE Transactions on Circuits and Systems for Video Technology},
  volume={33},
  number={12},
  pages={7842--7856},
  year={2023},
  publisher={IEEE}
}

@INPROCEEDINGS{ms_ssim,
  author={Wang, Z. and Simoncelli, E.P. and Bovik, A.C.},
  booktitle={The Thrity-Seventh Asilomar Conference on Signals, Systems and Computers, 2003}, 
  title={Multiscale structural similarity for image quality assessment}, 
  year={2003},
  volume={2},
  number={},
  pages={1398-1402 Vol.2},
  doi={10.1109/ACSSC.2003.1292216}}

@article{tsereh2024jpegai,
title={JPEG AI Image Compression Visual Artifacts: Detection Methods and Dataset},
author={Tsereh, Daria and Mirgaleev, Mark and Molodetskikh, Ivan and Kazantsev, Roman and Vatolin, Dmitriy},
journal={arXiv preprint arXiv:2411.06810},
year={2024}
}

@inproceedings{deng2009imagenet,
  title={Imagenet: A large-scale hierarchical image database},
  author={Deng, Jia and Dong, Wei and Socher, Richard and Li, Li-Jia and Li, Kai and Fei-Fei, Li},
  booktitle={2009 IEEE conference on computer vision and pattern recognition},
  pages={248--255},
  year={2009},
  organization={Ieee}
}

@inproceedings{lin2014microsoft,
  title={Microsoft coco: Common objects in context},
  author={Lin, Tsung-Yi and Maire, Michael and Belongie, Serge and Hays, James and Perona, Pietro and Ramanan, Deva and Doll{\'a}r, Piotr and Zitnick, C Lawrence},
  booktitle={European conference on computer vision},
  pages={740--755},
  year={2014},
  organization={Springer}
}

@inproceedings{Uhrig2017THREEDV,
  author = {Jonas Uhrig and Nick Schneider and Lukas Schneider and Uwe Franke and Thomas Brox and Andreas Geiger},
  title = {Sparsity Invariant CNNs},
  booktitle = {International Conference on 3D Vision (3DV)},
  year = {2017}
}

@inproceedings{he2016deep,
  title={Deep residual learning for image recognition},
  author={He, Kaiming and Zhang, Xiangyu and Ren, Shaoqing and Sun, Jian},
  booktitle={Proceedings of the IEEE conference on computer vision and pattern recognition},
  pages={770--778},
  year={2016}
}

@software{yolo11_ultralytics,
  author = {Glenn Jocher and Jing Qiu},
  title = {Ultralytics YOLO11},
  version = {11.0.0},
  year = {2024},
  url = {https://github.com/ultralytics/ultralytics},
  orcid = {0000-0001-5950-6979, 0000-0003-3783-7069},
  license = {AGPL-3.0}
}

@article{depth_anything_v2,
  title={Depth Anything V2},
  author={Yang, Lihe and Kang, Bingyi and Huang, Zilong and Zhao, Zhen and Xu, Xiaogang and Feng, Jiashi and Zhao, Hengshuang},
  journal={arXiv:2406.09414},
  year={2024}
}

@article{kuznetsova2020open,
  title={The open images dataset v4: Unified image classification, object detection, and visual relationship detection at scale},
  author={Kuznetsova, Alina and Rom, Hassan and Alldrin, Neil and Uijlings, Jasper and Krasin, Ivan and Pont-Tuset, Jordi and Kamali, Shahab and Popov, Stefan and Malloci, Matteo and Kolesnikov, Alexander and others},
  journal={International journal of computer vision},
  volume={128},
  number={7},
  pages={1956--1981},
  year={2020},
  publisher={Springer}
}

@inproceedings{nes,
  title     = {Black-box Adversarial Attacks with Limited Queries and Information},
  author    = {Ilyas, Andrew and Engstrom, Logan and Athalye, Anish and Lin, Jessy},
  booktitle = {Proceedings of the 35th International Conference on Machine Learning (ICML)},
  series    = {Proceedings of Machine Learning Research},
  volume    = {80},
  pages     = {2137--2146},
  year      = {2018},
  publisher = {PMLR},
  url       = {https://proceedings.mlr.press/v80/ilyas18a.html}
}

@inproceedings{squareattack,
  title     = {Square Attack: {A} Query-Efficient Black-Box Adversarial Attack via Random Search},
  author    = {Andriushchenko, Maksym and Croce, Francesco and Flammarion, Nicolas and Hein, Matthias},
  booktitle = {Computer Vision -- {ECCV} 2020 -- 16th European Conference, Glasgow, UK, August 23--28, 2020, Proceedings, Part {XXIII}},
  editor    = {Vedaldi, Andrea and Bischof, Horst and Brox, Thomas and Frahm, Jan{-}Michael},
  series    = {Lecture Notes in Computer Science},
  volume    = {12368},
  pages     = {484--501},
  year      = {2020},
  publisher = {Springer},
  doi       = {10.1007/978-3-030-58592-1\_29},
  url       = {https://doi.org/10.1007/978-3-030-58592-1\_29}
}

@inproceedings{mprnet,
    title={Multi-Stage Progressive Image Restoration},
    author={Syed Waqas Zamir and Aditya Arora and Salman Khan and Munawar Hayat
            and Fahad Shahbaz Khan and Ming-Hsuan Yang and Ling Shao},
    booktitle={CVPR},
    year={2021}
}

@inproceedings{
disco,
title={{DISCO}: Adversarial Defense with Local Implicit Functions},
author={Chih-Hui Ho and Nuno Vasconcelos},
booktitle={Advances in Neural Information Processing Systems},
editor={Alice H. Oh and Alekh Agarwal and Danielle Belgrave and Kyunghyun Cho},
year={2022},
url={https://openreview.net/forum?id=vgIz0emVTAd}
}

@misc{clic2025,
  title        = {CLIC 2025: 7th Challenge on Learned Image Compression},
  howpublished = {\url{https://clic2025.compression.cc/}},
  year         = {2025},
  note         = {Accessed: 2026-01-23}
}

@article{song2024training,
  title={A training-free defense framework for robust learned image compression},
  author={Song, Myungseo and Choi, Jinyoung and Han, Bohyung},
  journal={arXiv preprint arXiv:2401.11902},
  year={2024}
}

@article{kalmykov2026t,
  title={T-MLA: A targeted multiscale log--exponential attack framework for neural image compression},
  author={Kalmykov, Nikolay I and Dibo, Razan and Shen, Kaiyu and Zhonghan, Xu and Phan, Anh-Huy and Liu, Yipeng and Oseledets, Ivan},
  journal={Information Sciences},
  pages={123143},
  year={2026},
  publisher={Elsevier}
}
\bibliographystyle{icml2026}

%%%%%%%%%%%%%%%%%%%%%%%%%%%%%%%%%%%%%%%%%%%%%%%%%%%%%%%%%%%%%%%%%%%%%%%%%%%%%%%
%%%%%%%%%%%%%%%%%%%%%%%%%%%%%%%%%%%%%%%%%%%%%%%%%%%%%%%%%%%%%%%%%%%%%%%%%%%%%%%
% APPENDIX
%%%%%%%%%%%%%%%%%%%%%%%%%%%%%%%%%%%%%%%%%%%%%%%%%%%%%%%%%%%%%%%%%%%%%%%%%%%%%%%
%%%%%%%%%%%%%%%%%%%%%%%%%%%%%%%%%%%%%%%%%%%%%%%%%%%%%%%%%%%%%%%%%%%%%%%%%%%%%%%
\newpage
\appendix
\onecolumn
\section{Appendix}
\subsection{Contents}
Here, we briefly summarize the contents of all sections in this supplementary file:
\begin{itemize}
\item Section \ref{sec:related_works} lists and discusses the related works on NIC models;
\item Section \ref{sec:transferability} discusses the transferability of adversarial examples between NIC models;
\item Section \ref{sec:speed} provides computational performance measurements for all evaluated NIC models;
\item Section \ref{sec:wilcoxon} reports the results of statistical tests measuring the significance of codec comparisons;
\item Section \ref{sec:params} details the attack parameter search process and its impact on NIC performance; 
\item Section \ref{sec:delta_relation} analyzes the relationship between $\delta$ and $\Delta$ scores used throughout the study;
\item Section \ref{sec:defense_methodology} provides details on adversarial defenses implemented in NIC-RobustBench;
\item Section \ref{sec:cv_tasks} provides evaluation results for downstream computer vision tasks;
\item Section \ref{app:additional_results} reports additional results that complement the figures and insights mentioned in the main paper;
\item Section \ref{app:query_based_attacks} provides results for query-based black-box attacks on NIC models;
\item Section \ref{sec:repr} provides details on reproducibility; 
\item Section \ref{sec:limitations} discusses the limitations of this study and highlights the directions for future work;
\end{itemize}

\subsection{Related NIC works}
\label{sec:related_works}

\textbf{Neural image compression} has seen rapid progress in recent years. %Ballé et al.
\cite{balle2016end} introduced one of the first models for end-to-end compression with generalized divisive normalization and uniform scalar quantization. %Agustsson et al.
\cite{agustsson2017soft} proposed a soft-to-hard vector quantization method for compressive autoencoders. %Ballé et al.
\cite{balle2018variational} further modeled image compression as a variational autoencoder problem, introducing hyperpriors to improve entropy modeling. %Minnen et al.
\cite{minnen2018joint} extended hierarchical Gaussian scale mixture models with Gaussian mixtures and autoregressive components. %Mentzer et al.
\cite{mentzer2020high} utilized GANs to enhance perceptual quality in neural compression.
%Cheng et al.
\cite{cheng2020learned} developed an efficient entropy model based on discretized Gaussian mixtures and attention modules. %Yang et al.
\cite{nic1} introduced rate and complexity control using slimmable modules. %He et al.
\cite{he2022elic} proposed the ELIC model, which combines stacked residual blocks with a spatial-channel context entropy model. %Zou et al.
\cite{zou2022devil} enhanced compression with window-based attention, training both CNN and Transformer models. %Liu et al.
\cite{liu2023learned} combined transformer-CNN mixture blocks with a Swin-Transformer attention module. %Duan et al.
\cite{duan2023lossy} adopted a hierarchical VAE with a uniform posterior and a Gaussian-convolved uniform prior. %Wang et al.
\cite{wang2023evc} focused on real-time compression using residual and depth-wise convolution blocks, introducing mask decay and sparsity regularization for model distillation. %Yang et al.
\cite{yang2024lossy} presented a lossy compression scheme employing contextual latent variables and a diffusion model for reconstruction.
The novel JPEG AI standard \cite{jpeg_ai_standard} applies learned quantization across the entire image, surpassing traditional block-based methods in efficiency. It offers high and base operation points to balance compression efficiency and computational complexity, and includes adaptive tools depending on codec configuration.

\subsection{Transferability of adversarial examples}
\label{sec:transferability}

This section describes our experiments on the transferability of adversarial examples generated for one NIC model to another. The process of attacking NIC models using transferability can be described as follows: we attack one NIC model using a certain attack from our framework and feed the generated adversarial example to another NIC model. The attacks were performed using the ``Reconstruction Loss'' objective with two parameter presets that are averaged to get final results.

\begin{figure*}[!h]
\centering
\includegraphics[width=.99\textwidth]{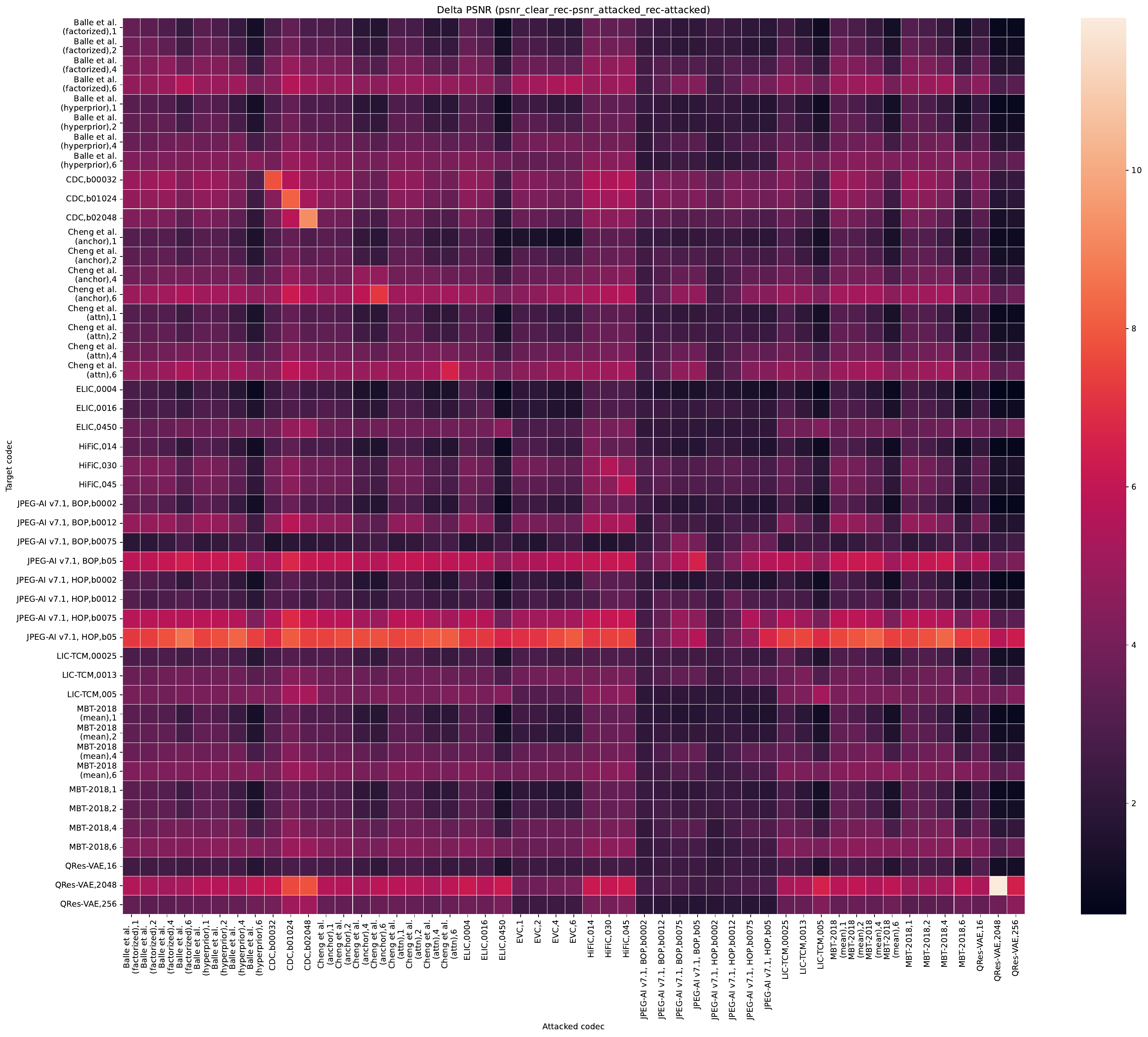}
\caption{Transferability of adversarial attacks constructed for codecs listed in columns to other codecs (listed in rows). The metric for attack success is $\Delta$ PSNR.}
\label{fig:transferability}
%\vspace*{-0.4cm}
\end{figure*}

Fig. \ref{fig:transferability} shows the transferability matrix with columns representing source codec versions on which adversarial examples are generated, and rows represent the target codec version to which the perturbations are transferred. We compare different bitrates of chosen codecs; they are included several times. 

First, we observe strong intra-model transferability, particularly across bitrates of specific versions of each NIC model, for example, for the CDC model. Perturbations from lower bitrates transferred to higher ones consistently remain effective, suggesting that transferability attacks exploit architecture vulnerabilities.

Second, the matrix concludes that there is also inter-model transferability. THE JPEG AI model demonstrates that there is transferability from a less stable NIC model to a more stable one. This indicates a strong dependence of transferability on architecture-specific features.

%Perturbations from lower bitrates transferred to higher ones show greater transferability than vice versa, which demonstrates a stronger effect of the successful attack on NIC with a lower bitrate. 

% The columns correspond to the source codec on which adversarial examples are generated, while each row represents the target codec to which the perturbations are transferred. We compare different bitrates of chosen codecs; they are included several times. The experiment employed Reconstruction Loss as the attack objective. We select two parameter presets for the attacks and average the results among them. Fig. \ref{fig:transferability} demonstrates high transferability between different bitrates of specific versions of each NIC model, especially from lower bitrates to higher ones. There is also transferability from a more stable NIC model to a less stable one.

% AR GB

\subsection{Computational complexity analysis}
\label{sec:speed}
\begin{figure*}[!h]
\centering
\includegraphics[width=.9\textwidth]{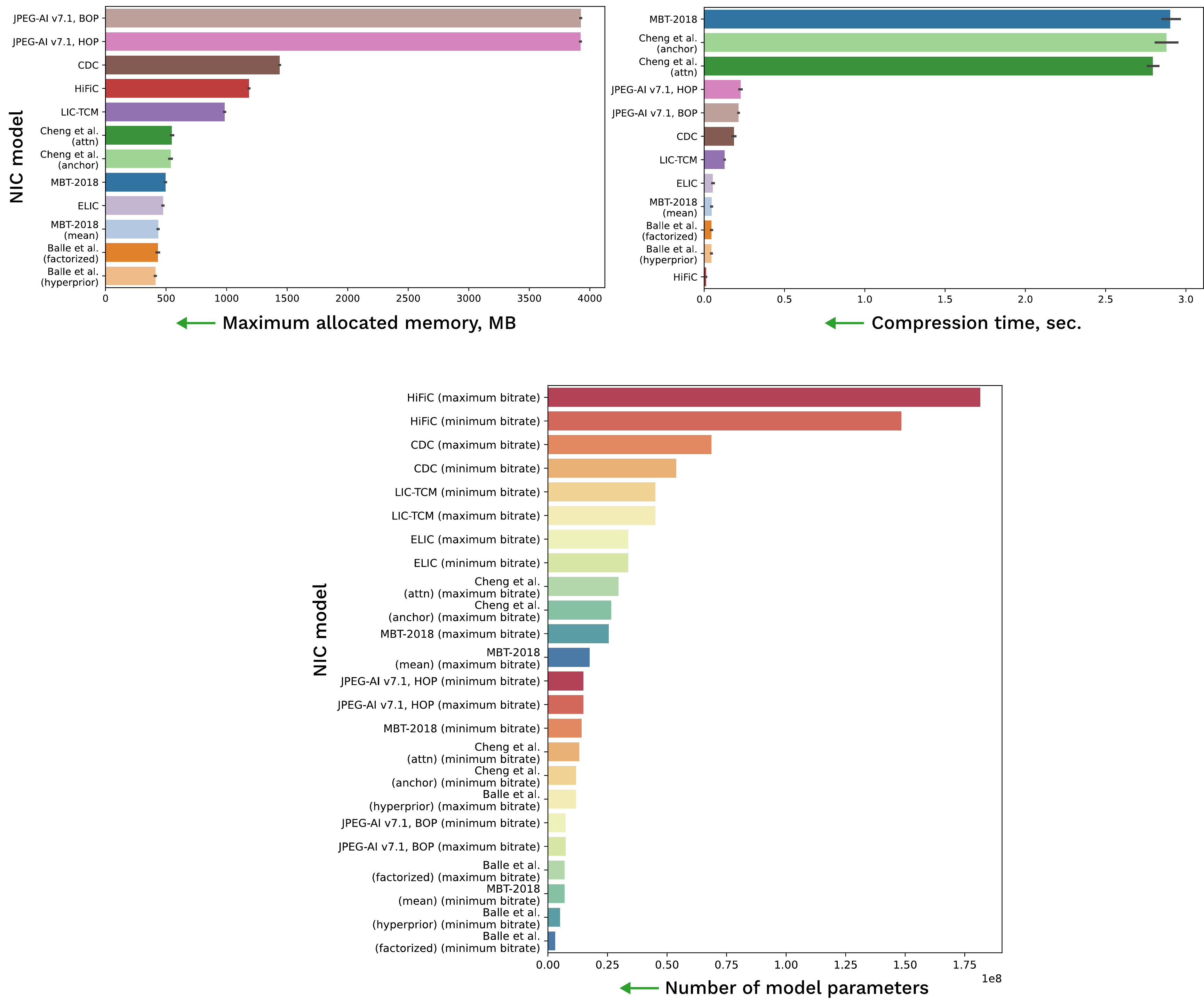}
\caption{Computational complexity of different NIC models measured in maximum allocated memory during the compression-decompression process (a), compression time (b), and number of model parameters (c).}
\label{fig:time_size_params}
%\vspace*{-0.4cm}
\end{figure*}

% AR EK
Fig. \ref{fig:time_size_params} compares the computational complexity of different NIC models. We highlight several insights from these performance evaluations: 
\begin{itemize}
    \item JPEG AI prioritizes accuracy/quality, but among the slowest and most memory-hungry in practice.
    \item Peak memory varies by $\sim\times10$. Several pipelines (e.g., JPEG AI BOP/HOP, CDC) peak around 3–4 GB, while classical baselines (factorized / hyperprior) stay well below 1 GB.
    \item HiFiC carries the largest parameter budget ($\sim10^7$) yet is among the fastest; conversely, several mid-sized models run slowly. Compute is dominated by architecture and operator choice (context models, attention/masked convs, entropy loops), not just parameter size.
    \item Memory and latency are correlated, but not monotonically. Attention/context-heavy designs generally incur both high memory and time; some GAN/hyperprior models decouple the two (large parameter numbers, low latency).
\end{itemize}

\newpage

\subsection{Statistical tests}
\label{sec:wilcoxon}
\begin{figure*}[!h]
\centering
\includegraphics[width=.99\textwidth]{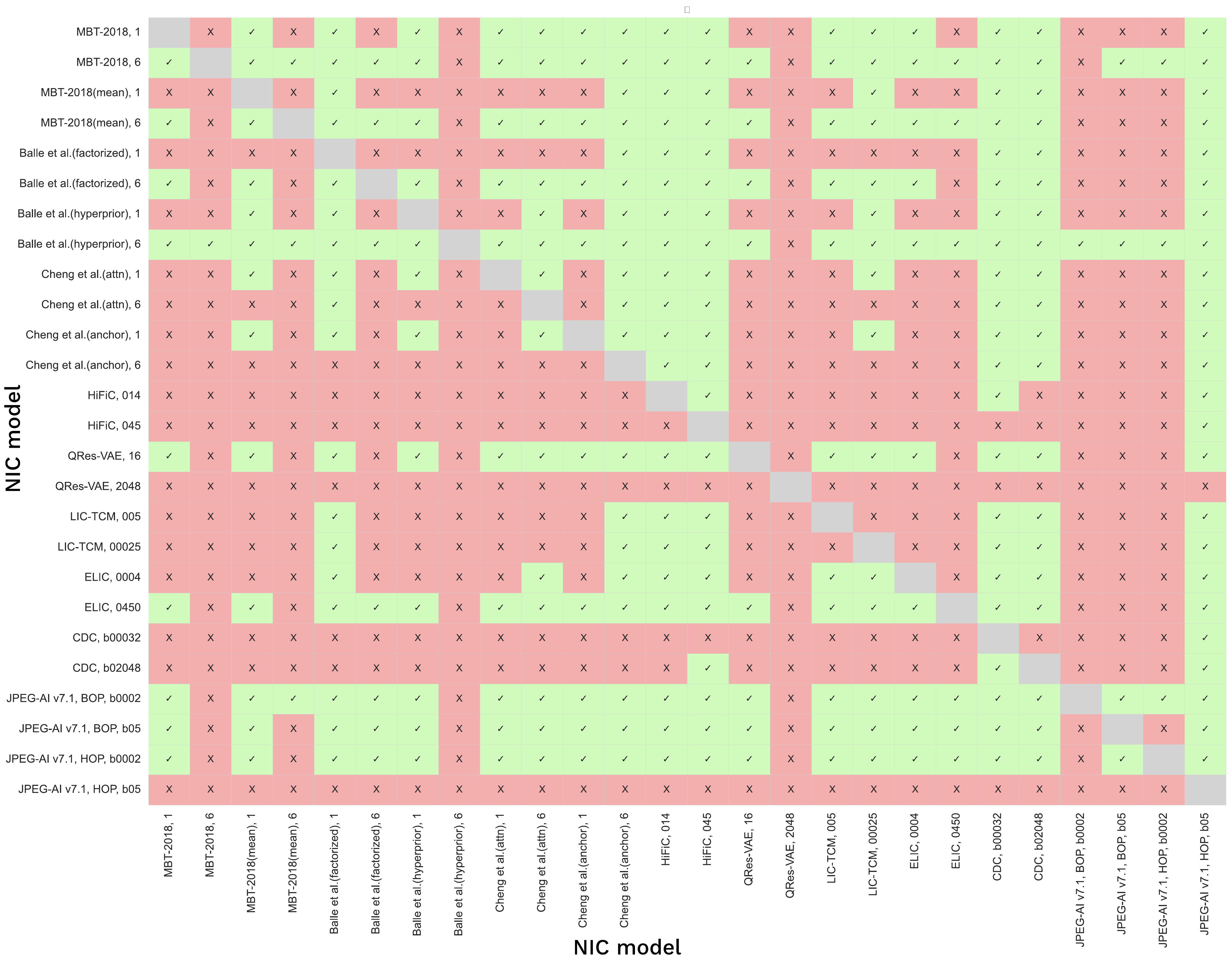}
\caption{Results of Wilcoxon signed-rank tests between different NICs based on \textbf{$\delta$SSIM} scores. $\checkmark$ symbols represent cases when the NIC model denoted in the row statistically outperforms the NIC defined in the column with a $p$-value$<0.05$. Bonferroni correction is used to account for a large number of pair-wise comparisons. In this experiment, we employ all attacks with the ``Reconstruction Loss'' objective.}
\label{fig:wilcoxon_delta_ssim}
%\vspace*{-0.4cm}
\end{figure*}

\begin{figure*}[!h]
\centering
\includegraphics[width=.99\textwidth]{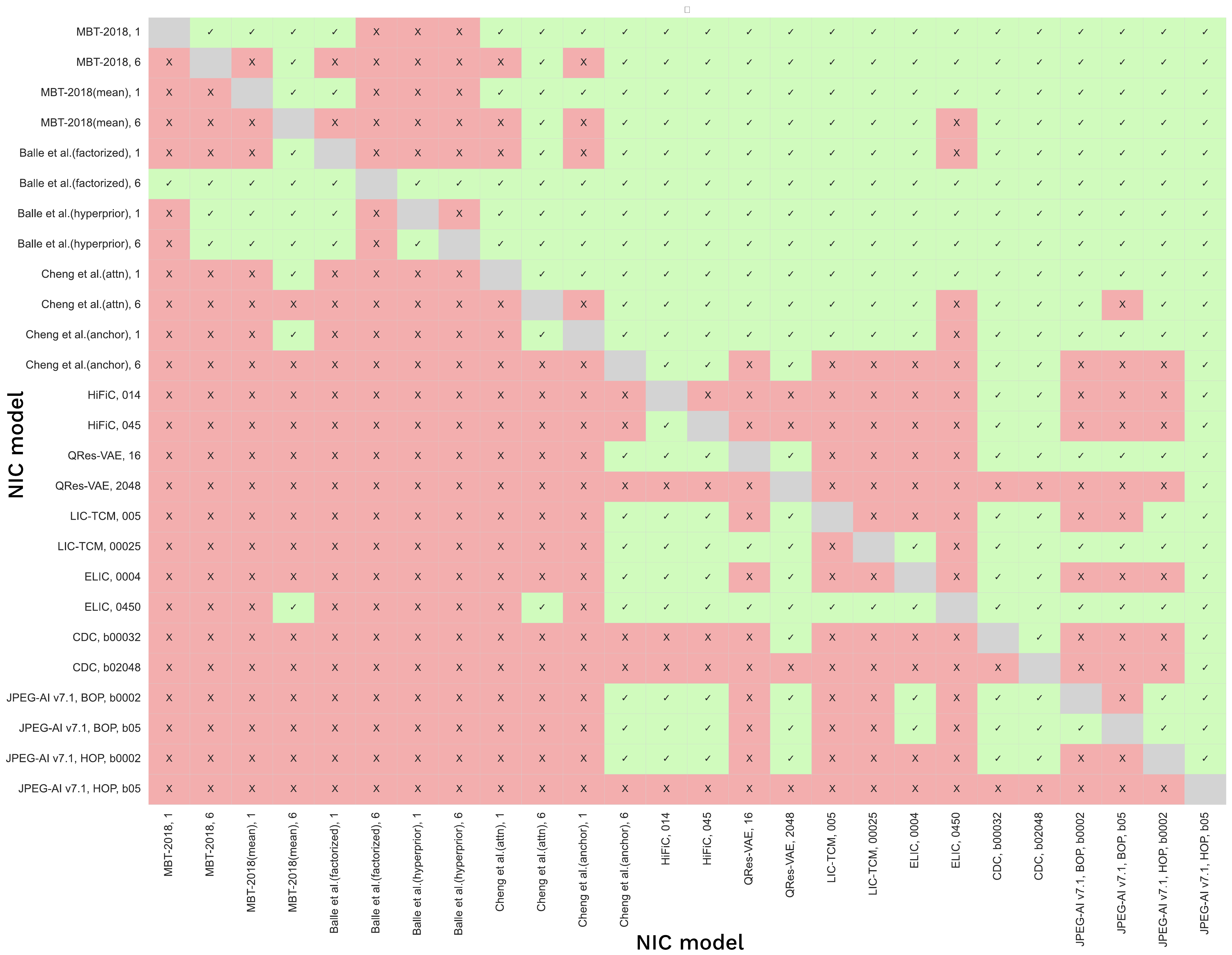}
\caption{Results of Wilcoxon signed-rank tests between different NICs based on \textbf{$\Delta$SSIM} scores. $\checkmark$ symbols represent cases when the NIC model denoted in the row statistically outperforms the NIC defined in the column with a $p$-value$<0.05$. Bonferroni correction is used to account for a large number of pair-wise comparisons. Here, we employ all attacks with the ``Reconstruction Loss'' objective.
}
\label{fig:wilcoxon_Delta_ssim}
%\vspace*{-0.4cm}
\end{figure*}

% AR AKh
We applied the one-sided Wilcoxon Signed Rank Test to assess the statistical significance of codec comparisons, as it is non-parametric and suited for paired samples without assuming normality (which restricts potential distributions of evaluation scores) --- ideal for adversarial robustness analysis. This test evaluates whether one NIC consistently outperforms another in terms of $\Delta$ scores and other relevant metrics. Results for $\Delta$SSIM and $\delta$SSIM scores are provided in Fig. \ref{fig:wilcoxon_Delta_ssim} and \ref{fig:wilcoxon_delta_ssim} respectively. To increase readability, we only use 2 compression ratios for each NIC model~(out of 4), the highest and lowest.

To ensure reliability, we applied the Bonferroni correction, which controls the family-wise error rate across hundreds of comparisons (13 codecs$\times$ 2 compression ratio per codec, squared). This conservative adjustment minimizes false positives, reinforcing the significance of the results.

\newpage

\subsection{Parameter sensitivity}
\label{sec:params}
\begin{figure*}[!h]
\centering
\includegraphics[width=.6\textwidth]{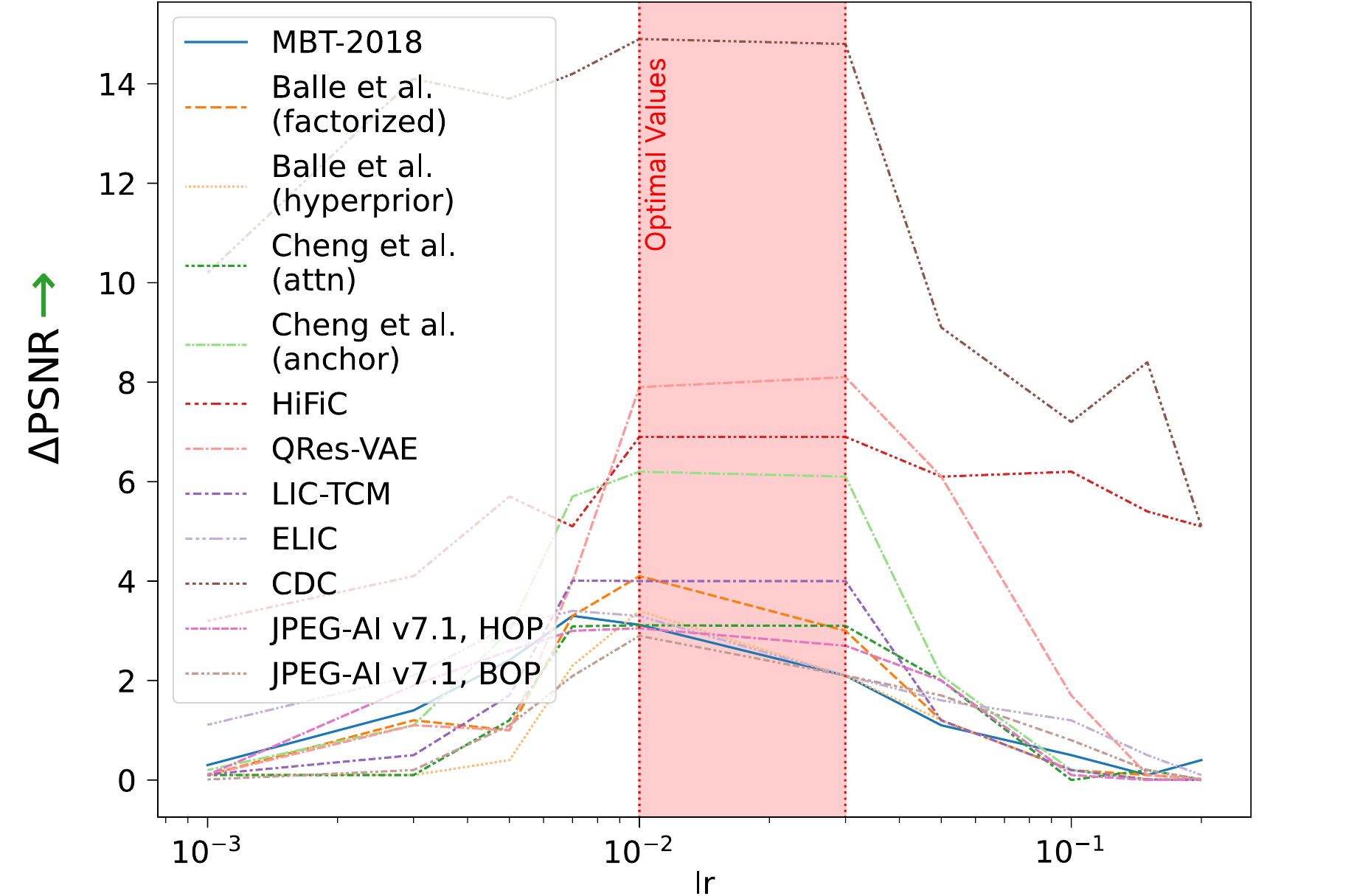}
\caption{Comparison of the effectiveness of MADC family attacks depending on the learning rate}
\label{fig:lr_madc}
%\vspace*{-0.4cm}
\end{figure*}

% AR EK
Fig. \ref{fig:lr_madc} demonstrates the relationship between attack effectiveness and the learning rate, a key hyperparameter determining the convergence rate of the attack. The effectiveness of attack methods can vary greatly depending on the specific attack parameters chosen for each codec. To ensure that the attacks function correctly, their parameters are selected based on a grid of potential values for each codec individually. Afterwards, presets (sets of attack parameters) are created that are suitable for a wide range of codecs. For example, attacks based on MADC with a learning rate parameter range from $\{lr=0.01\}$ to $\{lr=0.04\}$ proved to be a practical compromise to perform well across all codecs.

\newpage

\subsection{Relationship between $\delta$ and $\Delta$ robustness scores}
\label{sec:delta_relation}
\begin{figure*}[!h]
\centering
\includegraphics[width=.99\textwidth]{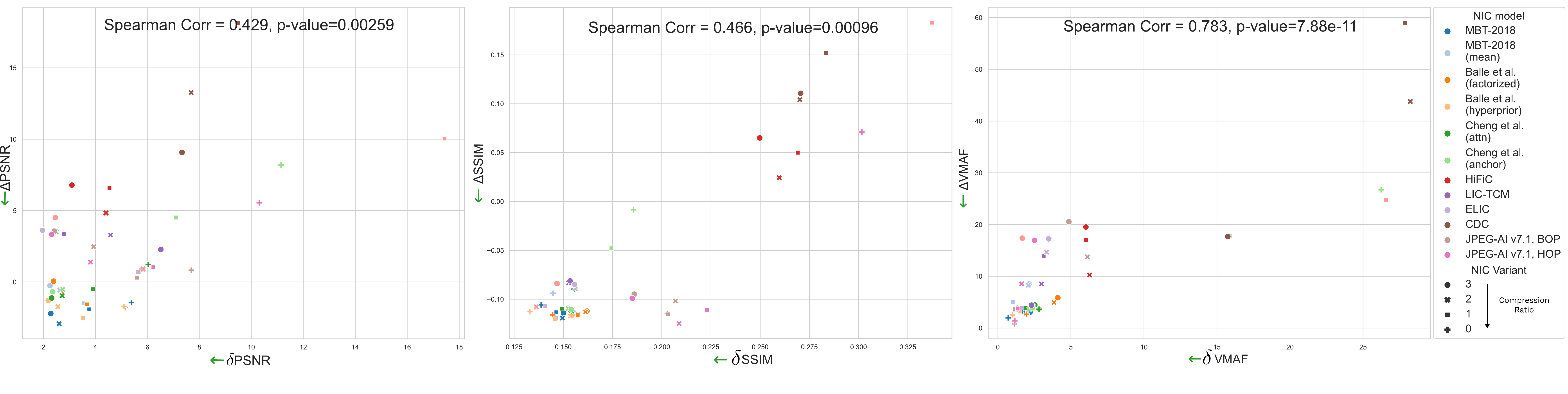}
\caption{Relationships between $\Delta$ and $\delta$ scores for different FR quality assessment metrics.}
\label{fig:corrs_deltas}
%\vspace*{-0.4cm}
\end{figure*}

% AR AKh
Fig.~\ref{fig:corrs_deltas} illustrates the relationship between different methods for calculating quality degradation after an attack. While $\Delta$ and $\delta$ are calculated between various pairs of clean, adversarial, decoded, and adversarial decoded images, their values generally correlate, and the lists of best- and worst-performing codecs are mostly similar between the two. The highest Spearman correlation between $\delta$ and $\Delta$ is 0.783 using VMAF as a quality metric. These results indicate that the proposed $\delta$ metric is a reasonable approach to measuring adversarial stability. The lower correlation for other full-reference (FR) metrics suggests that $\delta$ and $\Delta$ capture different aspects of the adversarial robustness of NIC models. 

\subsection{Adversarial defenses parameters}
\label{sec:defense_methodology}

Table \ref{tab:defence_details} presents a description of the adversarial defenses integrated into the framework, including their parameters for the forward and inverse transformations and time of application in the preprocess step. Note that neural-network-based methods require significantly more time for application than reversible transformations. We do not state the time required for the postprocessing step due to all of them being negligible (less than 1 ms overhead).

% \begin{table*}[!h]
% \centering
% \resizebox{\columnwidth}{!}{%
%     \begin{tabular}{lccccl}
%     Defense method & Type & Parameters & Preprocess ($Y = T(X)$) & Postprocess($X = T^{-1}(Y)$) \\
%     \hline
%     Flip & Spat. transf. & --- & $\text{flip}(X, [2, 3])$ & $\text{flip}(Y, [2, 3])$ \\
%     Random roll & Spat. transf. & \makecell[c]{$\text{dim} \in \{2, 3\}$, \\ $\text{size} \in [0, \text{len}(X[\text{dim}]) - 1]$} & $\text{roll}(X, \text{size}, \text{dim})$& $\text{roll}(Y, -\text{size}, \text{dim})$\\
%     Random rotate & Spat. transf. & $\theta \in [0, 359]$ & $\text{pad}(\text{rotate}(X, \theta))$& $\text{crop}(\text{rotate}(Y, -\theta))$  \\
%     \makecell[l]{Random color \\reorder} & Color transf. & $\sigma: \{0, 1, 2\} \to \{0, 1, 2\}$ & $X[:, \sigma([0, 1, 2])]$& $Y[:, \sigma^{-1}([0, 1, 2])]$\\
%     Random ens. & Ensemble & --- & Varies & Varies \\
%     \makecell[l]{Geometric \\self-ens. \cite{chen2023toward}} & Ensemble & --- & Varies & Varies \\
%     DiffPure \cite{nie2022diffusion} & Purification & --- & $\text{diffpure}(X)$ & $Y$ \\
%     \end{tabular}
% }
% \caption{List of adversarial defenses used in our paper. The }
% \label{tab:defence_details}
% \end{table*}

\begin{table}[!h]
\centering
\resizebox{\columnwidth}{!}{%
    \begin{tabular}{lccccc}
    Defense method & Type & Parameters & Preprocess ($Y = T(X)$) & Postprocess($X = T^{-1}(Y)$) & Time for preprocess (ms) \\
    \hline
    Flip & Spat. transf. & --- & $\text{flip}(X, [2, 3])$ & $\text{flip}(Y, [2, 3])$ & $0.81$ \\
    Random roll & Spat. transf. & \makecell[c]{$\text{dim} \in \{2, 3\}$, \\ $\text{size} \in [0, \text{len}(X[\text{dim}]) - 1]$} & $\text{roll}(X, \text{size}, \text{dim})$ & $\text{roll}(Y, -\text{size}, \text{dim})$ & $0.5$ \\
    Random rotate & Spat. transf. & $\theta \in [0, 359]$ & $\text{rotate}(\text{pad}(X), \theta)$ & $\text{crop}(\text{rotate}(Y, -\theta))$ & $4.21$ \\
    \makecell[l]{Random color \\reorder} & Color transf. & $\sigma: \{0, 1, 2\} \to \{0, 1, 2\}$ & $X[:, \sigma([0, 1, 2])]$ & $Y[:, \sigma^{-1}([0, 1, 2])]$ & $0.38$ \\
    Random ens. & Ensemble & --- & Varies & Varies & $0.49$ \\
    \makecell[l]{Geometric \\self-ens. \cite{chen2023toward}} & Ensemble & --- & Varies & Varies & $199.84$ \\
    DiffPure \cite{nie2022diffusion} & Purification & --- & $\text{diffpure}(X)$ & $Y$ & $306.75$ \\
    DISCO~\cite{disco} & Purification & --- & $\text{disco}(X)$ & $Y$ & $172.93$ \\
    MPRNet~\cite{mprnet} & Purification & --- & $\text{mrpnet}(X)$ & $Y$ & $52.26$ \\
    \end{tabular}
}
\caption{List of adversarial defenses used in our paper. The time was measured on the Kodak dataset size ($768  \times512$) on the Cheng et al. (attn) model.}
\label{tab:defence_details}
\end{table}

\newpage

\subsection{Attacks on downstream computer vision tasks}
\label{sec:cv_tasks}

Our paper aims to evaluate the impact on various computer vision (CV) tasks, notably classification, detection, and depth estimation. The evaluation can be described by the following idea: we feed the attacked image to the CV model, validate relevant metrics, and compare them with the original scores. For the given image $x \in X$ and CV model $F: X \to S$, where $S$ is the output of the CV model (logits in case of image classification, bounding box coordinates in case of detection and $\mathbb{R}^{H\times W}$ in case of depth estimation), we can describe the evaluation by the following equation:
\begin{equation}
    \begin{aligned}
    s= F\left(\underset{x': \rho(x',x) \le \varepsilon} {\arg\max} \: L(x, x', C(x), C(x'))\right)
    \end{aligned}
\end{equation}
where $x \in X$ is the original image, and $s \in S$ is the output of the CV model.

ImageNet \cite{deng2009imagenet} for classification, MS COCO \cite{lin2014microsoft} for detection, and KITTI Depth \cite{Uhrig2017THREEDV} for depth prediction are used for testing downstream tasks due to availability. The evaluation metrics used are common for the relevant tasks: accuracy for classification, precision, recall, and $F_1$ for classification and detection, IoU for detection, and MAE for depth estimation. 

We evaluate the mentioned metrics and demonstrate the difference between clean and attacked reconstructed images (denoted as $\Delta$ metrics) averaged by all attacks using the ``Reconstruction Loss'' objective in Table \ref{tab:cv_tasks}. We have taken ResNet50 \cite{he2016deep} for classification, YOLO11 \cite{yolo11_ultralytics} for detection, and Depth Anything V2 \cite{depth_anything_v2} for depth estimation. The results demonstrate that classification and depth estimation are robust to adversarial attacks on NIC models, with the gains not being noticeable. However, the task of detection is significantly impacted by adversarial attacks, resulting in a more significant metric difference. 

Adversarial examples from the ELIC model are the most efficient in attacking the classification model, while examples from QRes-VAE are efficient in attacking detection and depth estimation models.

\begin{table}[!h]
\resizebox{1.0\textwidth}{!}{%
\begin{tabular}{l|cccc|cccc|c}
\toprule
 & \multicolumn{4}{c}{Classification} & \multicolumn{4}{c}{Detection} & Depth Estimation \\
NIC & $\Delta$Accuracy & $\Delta$Precision & $\Delta$Recall & $\Delta$F1 & $\Delta$Precision & $\Delta$Recall & $\Delta$F1 & $\Delta$IoU & $\Delta$MAE \\
\midrule
Balle et al.(factorized), 1 & -0.020 & -0.024 & -0.019 & -0.023 & 0.018 & 0.019 & 0.019 & -0.004 & -0.001 \\
Balle et al.(factorized), 2 & 0.010 & 0.005 & 0.004 & 0.005 & 0.024 & 0.043 & 0.033 & -0.006 & -0.001 \\
Balle et al.(factorized), 4 & -0.030 & -0.036 & -0.037 & -0.036 & 0.001 & 0.004 & 0.002 & 0.007 & 0.002 \\
Balle et al.(factorized), 6 & -0.005 & -0.006 & -0.004 & -0.005 & 0.023 & 0.004 & 0.015 & -0.013 & 0.001 \\
Balle et al.(hyperprior), 1 & -0.005 & -0.007 & -0.002 & -0.005 & -0.034 & -0.034 & -0.031 & -0.012 & 0.000 \\
Balle et al.(hyperprior), 2 & -0.003 & -0.008 & -0.004 & -0.007 & -0.049 & -0.025 & -0.036 & -0.000 & -0.002 \\
Balle et al.(hyperprior), 6 & -0.005 & -0.004 & -0.004 & -0.004 & -0.006 & -0.006 & -0.007 & -0.004 & -0.001 \\
CDC, b00032 & 0.003 & 0.008 & 0.010 & 0.009 & 0.013 & -0.024 & -0.006 & -0.004 & 0.005 \\
CDC, b01024 & -0.017 & -0.026 & -0.023 & -0.025 & -0.014 & -0.019 & -0.016 & -0.037 & 0.003 \\
CDC, b02048 & -0.069 & -0.078 & -0.076 & -0.077 & -0.062 & -0.076 & -0.070 & -0.020 & 0.004 \\
Cheng et al.(anchor), 1 & -0.008 & -0.010 & -0.006 & -0.008 & -0.045 & -0.013 & -0.030 & 0.003 & 0.002 \\
Cheng et al.(anchor), 2 & 0.010 & 0.008 & 0.007 & 0.008 & -0.045 & -0.037 & -0.038 & 0.005 & 0.002 \\
Cheng et al.(anchor), 4 & -0.030 & -0.030 & -0.031 & -0.031 & -0.042 & -0.037 & -0.038 & -0.047 & 0.006 \\
Cheng et al.(anchor), 6 & -0.057 & -0.054 & -0.052 & -0.053 & -0.085 & -0.073 & -0.081 & -0.073 & \underline{0.017} \\
Cheng et al.(attn), 1 & -0.013 & -0.009 & -0.009 & -0.009 & -0.034 & -0.009 & -0.021 & \underline{0.007} & -0.001 \\
Cheng et al.(attn), 2 & -0.020 & -0.024 & -0.019 & -0.022 & -0.002 & -0.021 & -0.016 & -0.004 & 0.000 \\
Cheng et al.(attn), 4 & 0.000 & -0.000 & -0.002 & -0.001 & -0.014 & -0.002 & -0.007 & -0.004 & 0.004 \\
Cheng et al.(attn), 6 & -0.026 & -0.032 & -0.034 & -0.033 & 0.010 & 0.006 & 0.008 & -0.009 & 0.002 \\
ELIC, 0004 & \textbf{0.030} & 0.024 & \underline{0.028} & 0.025 & -0.022 & -0.013 & -0.020 & -0.040 & 0.002 \\
ELIC, 0016 & -0.040 & -0.023 & -0.026 & -0.024 & \underline{0.046} & \underline{0.048} & 0.043 & -0.016 & 0.005 \\
ELIC, 0450 & 0.020 & \underline{0.027} & \textbf{0.029} & \textbf{0.028} & 0.013 & 0.012 & 0.014 & -0.001 & 0.002 \\
HiFiC, 014 & -0.061 & -0.060 & -0.054 & -0.058 & -0.073 & -0.064 & -0.065 & -0.161 & -0.002 \\
HiFiC, 030 & -0.012 & -0.012 & -0.012 & -0.012 & -0.025 & -0.020 & -0.023 & -0.022 & 0.001 \\
HiFiC, 045 & -0.020 & -0.023 & -0.024 & -0.023 & 0.009 & -0.006 & -0.000 & -0.003 & 0.001 \\
LIC-TCM, 00025 & -0.044 & -0.044 & -0.044 & -0.044 & -0.026 & -0.011 & -0.016 & -0.009 & 0.002 \\
LIC-TCM, 0013 & -0.013 & -0.018 & -0.019 & -0.018 & \textbf{0.049} & 0.024 & 0.036 & -0.015 & 0.005 \\
LIC-TCM, 005 & \underline{0.021} & \textbf{0.027} & 0.023 & \underline{0.026} & -0.067 & -0.075 & -0.071 & -0.010 & -0.003 \\
MBT-2018, 1 & 0.007 & 0.001 & 0.008 & 0.003 & 0.005 & -0.006 & 0.000 & -0.004 & 0.001 \\
MBT-2018, 2 & -0.030 & -0.031 & -0.029 & -0.030 & 0.012 & -0.006 & 0.005 & \textbf{0.008} & -0.002 \\
MBT-2018, 4 & 0.000 & 0.001 & 0.003 & 0.001 & 0.018 & 0.021 & 0.021 & -0.013 & -0.004 \\
MBT-2018, 6 & 0.003 & 0.008 & 0.008 & 0.008 & 0.007 & 0.008 & 0.008 & -0.004 & -0.000 \\
MBT-2018(mean), 1 & -0.035 & -0.034 & -0.036 & -0.035 & -0.017 & -0.019 & -0.018 & -0.002 & 0.000 \\
MBT-2018(mean), 2 & 0.006 & 0.012 & 0.010 & 0.011 & 0.032 & 0.003 & 0.016 & -0.007 & 0.001 \\
MBT-2018(mean), 4 & 0.006 & 0.008 & 0.007 & 0.008 & -0.038 & -0.001 & -0.019 & 0.003 & -0.002 \\
MBT-2018(mean), 6 & 0.006 & 0.011 & 0.013 & 0.012 & -0.018 & -0.000 & -0.008 & -0.002 & -0.001 \\
QRes-VAE, 16 & -0.020 & -0.026 & -0.020 & -0.024 & -0.022 & \textbf{0.094} & \underline{0.044} & -0.002 & 0.007 \\
QRes-VAE, 2048 & -0.187 & -0.152 & -0.149 & -0.151 & -0.157 & -0.188 & -0.173 & -0.195 & \textbf{0.035} \\
QRes-VAE, 256 & 0.010 & 0.015 & 0.019 & 0.016 & 0.041 & 0.047 & \textbf{0.045} & -0.007 & 0.006 \\
\bottomrule
\end{tabular}

}
\caption{Comparison of results under adversarial scenarios after compression for downstream tasks (image classification, object detection, and depth estimation).}
\label{tab:cv_tasks}
\end{table}

\newpage

\subsection{Additional results}
\label{app:additional_results}
\begin{figure*}[t!]
\centering
\includegraphics[width=.99\textwidth]{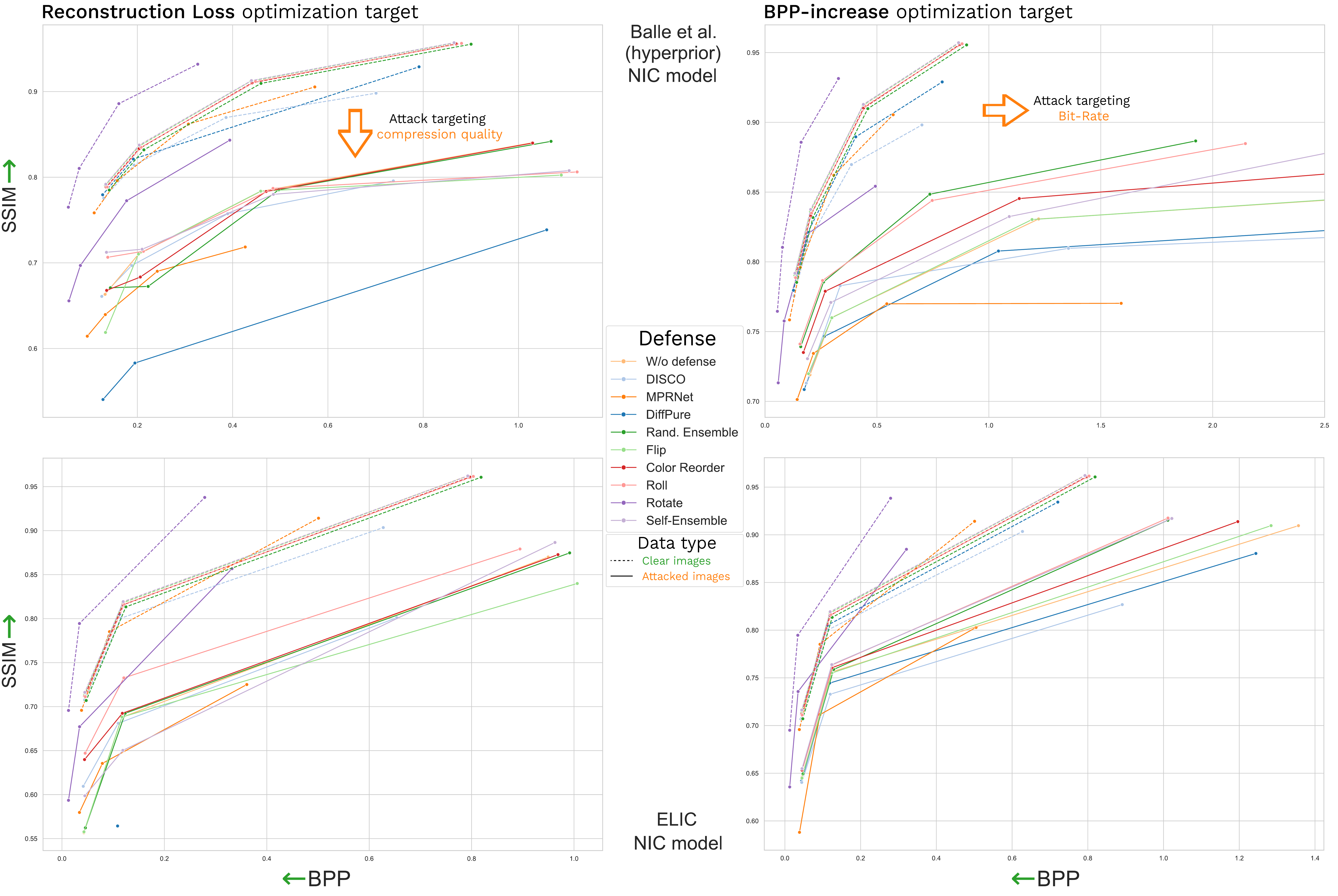}
\caption{RD-curves representing NIC performance on clear~(dashed lines) and attacked~(solid) data with different defenses as image preprocessing. The first row represents results for the Balle et al. (hyperprior) NIC model, and the second row --- for the ELIC model. Attacks target image quality on the left subfigure, and bitrate on the right.}
\label{fig:lineplots_defenses}
%\vspace*{-0.4cm}
\end{figure*}

\begin{figure*}[t!]
\centering
\includegraphics[width=.99\textwidth]{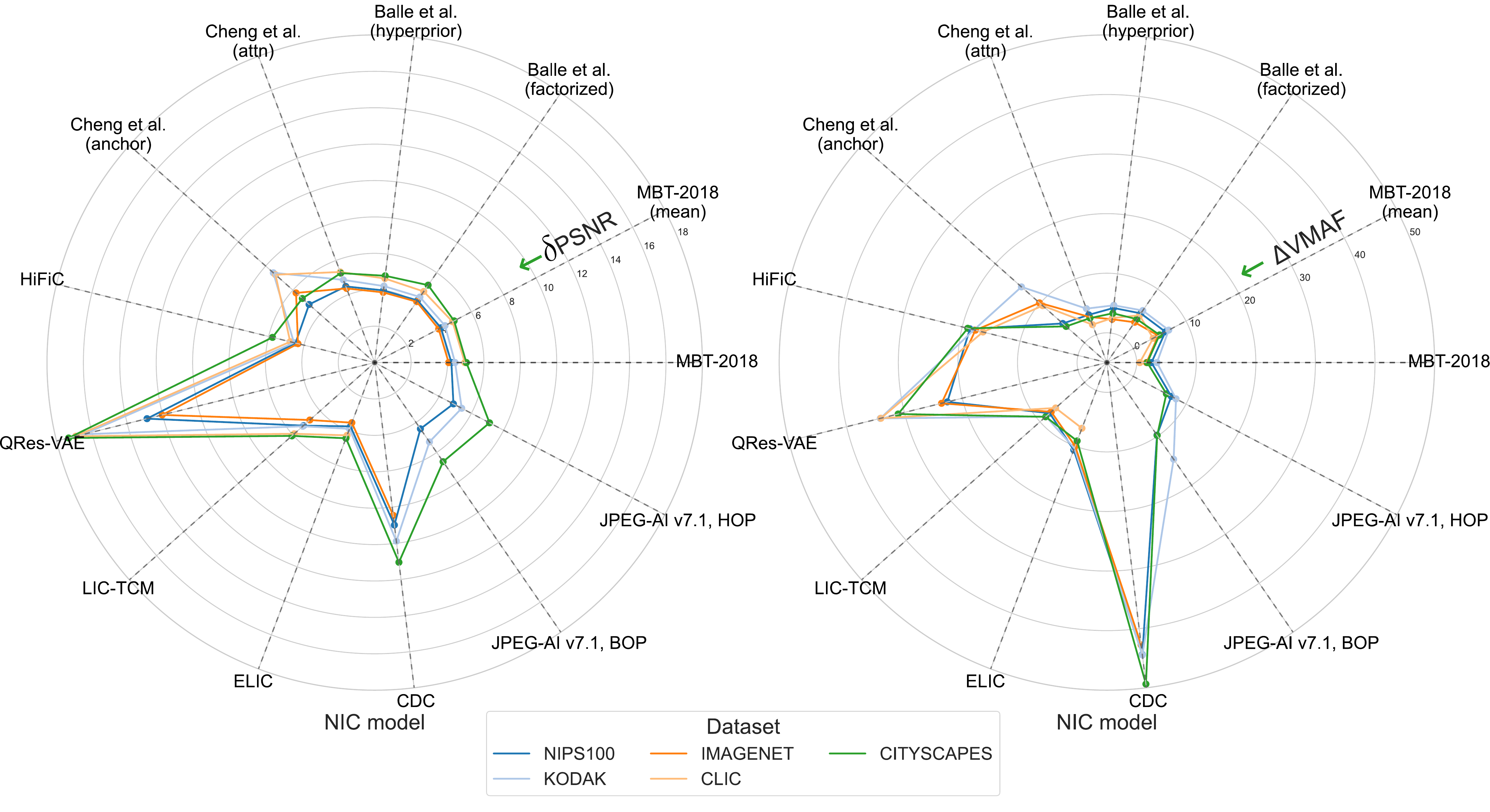}
\caption{NIC performance across different datasets measured with $\delta PSNR\downarrow$ and $\Delta VMAF \downarrow$. }
\label{fig:spiderplot_datasets}
%\vspace*{-0.4cm}
\end{figure*}

\begin{figure*}[t!]
\centering
\includegraphics[width=.99\textwidth]{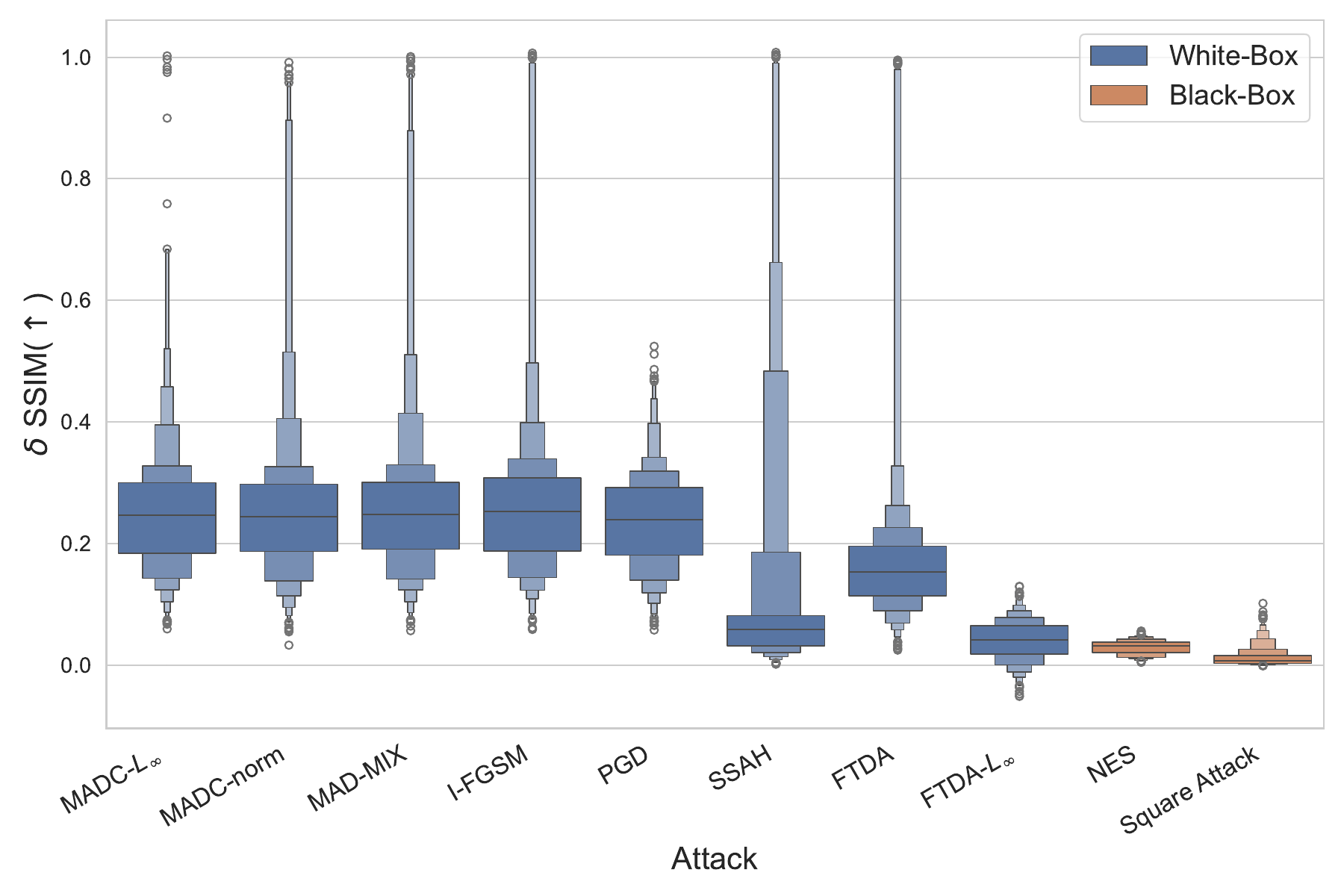}
\caption{NIC performance across different types of adversarial attacks measured with $\delta SSIM\uparrow$.}
\label{fig:attack_types}
%\vspace*{-0.4cm}
\end{figure*}

In this section, we provide some additional detailed results.

Table \ref{tab:by_losses_vmaf_aggregated} demonstrates $\Delta$VMAF and $\delta$VMAF scores across different attack objectives, with scores aggregated across all attacks and codec variants (i.e., different compression ratios of the same codec). Table \ref{tab:by_losses_vmaf} details these results and compares all NIC variants across different attack objectives. Table \ref{tab:by_losses_ssim} provides similar results for $\Delta$SSIM and $\delta$SSIM metrics.

% Table \ref{tab:by_losses_ssim} compares NIC models across different attacks objectives and reports $\Delta$SSIM and $\delta$SSIM metrics. TODO

Tables \ref{tab:by_attack_ssim} and \ref{tab:by_attack_psnr}  compare NIC models across different adversarial attacks with the same optimization target.

% Table \ref{tab:by_attack_psnr} compares NIC models across different attack objectives and TODO.

Table \ref{tab:metrics_all_averaged} summarizes all evaluation scores across all attacks with the ``Reconstruction Loss'' objective. The results are similar to the results from the main paper, demonstrating the generalization of our findings across the chosen datasets.

Fig. \ref{fig:lineplots_defenses} presents RD curves for SSIM across all attacks with the ``Reconstruction Loss'' and BPP loss objectives for all evaluated defenses and the ELIC model. The results demonstrate that most defenses alter the original curve (without defense) on clean images. Most defenses improve the SSIM between the input and reconstructed images for the attacked images, demonstrating the efficiency of the defenses. The most efficient defenses are Rotate, Self-ensemble, and DiffPure.

Fig. \ref{fig:spiderplot_datasets} summarizes $\delta PSNR$ and $\Delta VMAF$ across different datasets with the ``Reconstruction Loss'' objective.

\begin{table}[!h]
\resizebox{1.0\textwidth}{!}{%
\begin{tabular}{c|cc|cc|cc|cc|cc|cc}
\toprule
                 NIC & \multicolumn{2}{c}{Rec. Loss} & \multicolumn{2}{c}{BPP Loss} & \multicolumn{2}{c}{FTDA Loss} & \multicolumn{2}{c}{Added Noises} & \multicolumn{2}{c}{Rec. Loss(Y)} & \multicolumn{2}{c}{Source Rec.(Y)} \\
                            & $\Delta$VMAF$\downarrow$ & $\delta$VMAF$\downarrow$ & $\Delta$VMAF$\downarrow$ & $\delta$VMAF$\downarrow$ & $\Delta$VMAF$\downarrow$ & $\delta$VMAF$\downarrow$ & $\Delta$VMAF$\downarrow$ & $\delta$VMAF$\downarrow$ & $\Delta$VMAF$\downarrow$ & $\delta$VMAF$\downarrow$ & $\Delta$VMAF$\downarrow$ & $\delta$VMAF$\downarrow$ \\
\midrule
                   MBT-2018 &                \textbf{2.76} &                \textbf{1.61} &             \underline{6.26} &                         3.03 &                 \textbf{6.9} &                         1.17 &                \textbf{4.94} &                         1.66 &                \textbf{2.06} &             \underline{1.46} &                         7.37 &                        0.511 \\
 Balle et al. (hyperprior) &             \underline{3.61} &                         1.88 &                         8.82 &                         6.05 &             \underline{8.53} &                         1.46 &             \underline{6.83} &                         1.88 &                          2.7 &                          1.9 &                         8.85 &                          1.4 \\
       Cheng et al. (attn) &                         4.03 &                         2.43 &                          7.6 &                         5.01 &                         8.78 &                         2.06 &                         7.06 &                         2.63 &                         3.53 &                         2.98 &                         8.24 &                         1.26 \\
\bottomrule
\end{tabular}
}
\caption{$\Delta$VMAF ($\downarrow$) and $\delta$VMAF ($\downarrow$) scores for different NIC models across different attack objectives. Results are averaged across all attacks and all NIC variants.}
\label{tab:by_losses_vmaf_aggregated}
\end{table}

\begin{table}[!h]
\resizebox{1.0\textwidth}{!}{%
\begin{tabular}{cc|cc|cc|cc|cc|cc|cc}
\toprule
    &        & \multicolumn{2}{c}{Rec. Loss} & \multicolumn{2}{c}{BPP Loss} & \multicolumn{2}{c}{FTDA Loss} & \multicolumn{2}{c}{Added Noises} & \multicolumn{2}{c}{Rec. Loss(Y)} & \multicolumn{2}{c}{Source Rec.(Y)} \\
    &        &                       $\Delta$VMAF$\downarrow$ &                       $\delta$VMAF$\downarrow$ &                       $\Delta$VMAF$\downarrow$ &                       $\delta$VMAF$\downarrow$ &                     $\Delta$VMAF$\downarrow$ &                       $\delta$VMAF$\downarrow$ &                       $\Delta$VMAF$\downarrow$ &                       $\delta$VMAF$\downarrow$ &                       $\Delta$VMAF$\downarrow$ &                       $\delta$VMAF$\downarrow$ &                       $\Delta$VMAF$\downarrow$ &                       $\delta$VMAF$\downarrow$ \\
NIC & NIC Var. &                                                    &                                                    &                                                    &                                                    &                                                  &                                                    &                                                    &                                                    &                                                    &                                                    &                                                    &                                                    \\
\midrule
MBT-2018 & 1 &                        3.065 &                        2.206 &                       10.453 &                        4.027 &                       11.271 &                        1.212 &                        7.286 &                        2.018 &                        2.083 &                        2.594 &                       13.281 &                        0.373 \\
\bottomrule
\end{tabular}
}
\caption{$\Delta$VMAF ($\downarrow$) and $\delta$VMAF ($\downarrow$) scores for different NIC models and their variants across different attack objectives. Results are averaged across all attacks.}
\label{tab:by_losses_vmaf}
\end{table}

\begin{table}[!h]
\resizebox{1.0\textwidth}{!}{%
\begin{tabular}{cc|cc|cc|cc|cc|cc|cc}
\toprule
    &        & \multicolumn{2}{c}{Rec. Loss} & \multicolumn{2}{c}{BPP Loss} & \multicolumn{2}{c}{FTDA Loss} & \multicolumn{2}{c}{Added Noises} & \multicolumn{2}{c}{Rec. Loss(Y)} & \multicolumn{2}{c}{Source Rec.(Y)} \\
    &        &                         $\Delta$SSIM$\downarrow$ &                        $\delta$SSIM$\downarrow$ &                          $\Delta$SSIM$\downarrow$ &                          $\delta$SSIM$\downarrow$ &                          $\Delta$SSIM$\downarrow$ &                         $\delta$SSIM$\downarrow$ &                          $\Delta$SSIM$\downarrow$ &                         $\delta$SSIM$\downarrow$ &                         $\Delta$SSIM$\downarrow$ &                         $\delta$SSIM$\downarrow$ &                           $\Delta$SSIM$\downarrow$ &                          $\delta$SSIM$\downarrow$ \\
NIC & NIC Var. &                                                      &                                                     &                                                       &                                                       &                                                       &                                                      &                                                       &                                                      &                                                      &                                                      &                                                        &                                                       \\
\midrule
Balle et al. (hyperprior) & 1 &                       -0.116 &                        0.154 &                       -0.034 &                        0.094 &                       -0.007 &                        0.036 &                       -0.105 &                        0.110 &           \underline{-0.141} &                        0.195 &                       -0.011 &                        0.077 \\
\bottomrule
\end{tabular}
}
\caption{$\Delta$SSIM ($\downarrow$) and $\delta$SSIM ($\downarrow$) scores for different NIC models and their variants across different attack objectives. Results are averaged across all attacks.}
\label{tab:by_losses_ssim}
\end{table}

\begin{table}[!h]
\resizebox{1.0\textwidth}{!}{%
\begin{tabular}{lllllllllll}
\toprule
    &        &                                                 FTDA &                                  FTDA-$L_{{\infty}}$ &                                               I-FGSM &                                              MAD-MIX &                                                MADC &                                  MADC-$L_{{\infty}}$ &                                            MADC-norm &                                                  PGD &                                                 SSAH \\
NIC & NIC Var. &                                                      &                                                      &                                                      &                                                      &                                                     &                                                      &                                                      &                                                      &                                                      \\
\midrule
Balle et al. (hyperprior) & 1 &              \makecell{-0.094 \footnotesize{±0.048}} &              \makecell{-0.068 \footnotesize{±0.032}} &              \makecell{-0.155 \footnotesize{±0.064}} &              \makecell{-0.164 \footnotesize{±0.068}} &              \makecell{0.001 \footnotesize{±0.000}} &              \makecell{-0.163 \footnotesize{±0.068}} &              \makecell{-0.165 \footnotesize{±0.069}} &              \makecell{-0.176 \footnotesize{±0.072}} &              \makecell{-0.087 \footnotesize{±0.043}} \\
\bottomrule
\end{tabular}
}
\caption{$\Delta$SSIM ($\downarrow$) scores for different NIC models and their variants across different adversarial attacks. The ``Reconstruction Loss'' objective is used for all attacks.}
\label{tab:by_attack_ssim}
\end{table}

\begin{table}[!h]
\resizebox{1.0\textwidth}{!}{%
\begin{tabular}{lcccccccccc}
\toprule
    &        &                                                 FTDA &                                  FTDA-$L_{{\infty}}$ &                                               I-FGSM &                                              MAD-MIX &                                                 MADC &                                  MADC-$L_{{\infty}}$ &                                            MADC-norm &                                                  PGD &                                                 SSAH \\
NIC & NIC Var. &                                                      &                                                      &                                                      &                                                      &                                                      &                                                      &                                                      &                                                      &                                                      \\
\midrule
MBT-2018 & 1 &  \makecell{\underline{-4.770} \footnotesize{±2.040}} &              \makecell{-3.131 \footnotesize{±1.817}} &              \makecell{-5.168 \footnotesize{±1.518}} &              \makecell{-5.773 \footnotesize{±1.755}} &              \makecell{13.970 \footnotesize{±2.596}} &              \makecell{-5.737 \footnotesize{±1.731}} &              \makecell{-5.867 \footnotesize{±1.840}} &              \makecell{-6.320 \footnotesize{±1.938}} &              \makecell{-5.470 \footnotesize{±2.945}} \\
\bottomrule
\end{tabular}
}
\caption{$\Delta$PSNR ($\downarrow$) scores for different NIC models and their variants across different adversarial attacks. The ``Reconstruction Loss'' objective is used for all attacks.}
\label{tab:by_attack_psnr}
\end{table}

\begin{table}[t]
\resizebox{1.0\textwidth}{!}{%
\begin{tabular}{cc|ccccccccc}
\toprule
NIC & NIC Var. & $\Delta$SSIM$\downarrow$ & $\Delta$PSNR$\downarrow$ & $\Delta$MSSSIM$\downarrow$ & $\Delta$VMAF$\downarrow$ & $\delta$SSIM$\downarrow$ & $\delta$PSNR$\downarrow$ & $\delta$MSSSIM$\downarrow$ & $\delta$VMAF$\downarrow$ & $\Delta$BPP$\downarrow$ \\

\midrule
Balle et al. (hyperprior) & 1 &                       -0.117 &                         -1.3 &                         -0.012 &                         4.26 &                        0.155 &             \underline{2.19} &                          0.036 &                         2.52 &                 0.00069 \\
    & 2 &                       -0.117 &                        -1.74 &                         -0.013 &                         4.42 &                        0.154 &                         2.56 &                          0.032 &                         2.51 &                  0.0029 \\
    & 4 &           \underline{-0.120} &             \underline{-2.5} &                         -0.015 &                          3.2 &                        0.146 &                         3.53 &                          0.025 &                         1.49 &                   0.027 \\
    & 6 &                       -0.113 &                        -1.72 &                         -0.012 &                         2.55 &               \textbf{0.133} &                          5.1 &                 \textbf{0.019} &             \underline{1.01} &                    0.15 \\
Balle et al. (factorized) & 1 &                       -0.112 &                       0.0572 &                         -0.009 &                         5.87 &                        0.162 &                         2.39 &                          0.040 &                         4.11 &                 0.00069 \\
\bottomrule
\end{tabular}

}
\caption{Performance of NIC models across different evaluation metrics averaged across all attacks. The ``Reconstruction Loss'' objective is used as a target for all attacks.}
\label{tab:metrics_all_averaged}
\end{table}

\subsection{Query-based black-box attacks on NIC}
\label{app:query_based_attacks}
To our knowledge, there are no specialized black-box attacks targeting NIC. However, we understand the importance of the robustness evaluation to this type of attack. To evaluate the efficiency of black-box attacks, we have added and evaluated popular query-based attacks that include NES~\cite{nes} and Square Attack~\cite{squareattack}. We have changed the target loss function and used a similar procedure for white-box attacks to select the parameters of the methods (both of them are run with 10,000 queries per image). Fig. \ref{fig:attack_types} compares different attacks and attack types with the ``Reconstruction Loss'' objective. We conclude that these attacks are inefficient for the NIC task; thus, the question of developing an efficient query-based black-box attack on NIC remains open.

\subsection{Reproducibility}
\label{sec:repr}
% про открытый код
We will attach the full code for the experiments, pre-trained checkpoints, and a Docker image to the supplementary material and to a public repository upon acceptance.

All stages of the pipeline are easily configurable and customizable, as core run parameters can be specified via a single YAML configuration file. By leveraging Docker-based containerization and simple YAML-based setup configuration, NIC-RobustBench aims for experiment reproducibility. Additionally, NIC-RobustBench includes tools for results visualization, enabling researchers to obtain a visual summary of the experiments quickly.

We used a sophisticated end-to-end automated evaluation pipeline using Docker-based containerization and Slurm to ensure all our results are reproducible. All calculations required approximately 25,000 GPU-hours. Timing benchmarks were performed on a dedicated server with NVIDIA Tesla A100 80 Gb GPU, Intel Xeon Processor (Ice Lake) 32-Core Processor @ 2.60 GHz.

The high computational complexity of the benchmarking can be explained by the large evaluation grid: applying 6 attack algorithms with 6 adversarial objectives to target 9 NICs (more accurately, $>40$ variants considering different bitrates) will result in $>324$ images produced from the single clear example ($>1400$ considering different bitrates).

\subsection{Limitations}
\label{sec:limitations}
Our library and benchmark for attacking NIC is a contribution to NIC adversarial robustness. However, this paper has the following limitations:

\begin{itemize}
    \item Analysis is primarily empirical, and while we provide explanations for observed trends, the paper does not include a formal theoretical analysis of NIC robustness. Some experiments in our work include theoretical analysis (eg, Section \ref{sec:wilcoxon}, experiment with NIC model families). We recognize the importance of the theoretical analysis of the NIC robustness, but we consider extensive analysis as future work. Moreover, this can be significantly simplified with the use of our software.
    \item Our paper does not cover the case of adversarial defense stacking or combinations of adversarial defenses. This was omitted due to the computational complexity of the entire evaluation. This issue is partially covered by the evaluation of ensembling defenses.
    \item The computational cost of evaluating memory-consuming NIC models (e.g., JPEG AI, LIC-TCM) under some attacks can be significant, which may limit accessibility and reproducibility for researchers without specialized hardware (a GPU with enough memory capacity).
\end{itemize}

\iffalse
\begin{table}[t]
\caption{List of adversarial attacks used in our benchmark. WB and BB are white-box and black-box attack types.}
\begin{center}
    \resizebox{\linewidth}{!}{
    \begin{tabular}{lccccl}
    Adversarial attack & Type & Restriction & Preset 1 & Preset 2 \\
    \hline
    
    \end{tabular}
    }
\end{center}
\label{tab:attacks_details}
\end{table}
\fi

\end{document}